\documentclass[11pt]{article}

\usepackage[english]{babel}
\usepackage[utf8]{inputenc}
\usepackage[T1]{fontenc}
\usepackage{mathrsfs}

\usepackage[leqno]{amsmath}
\usepackage{color}

\definecolor{dukeblue}{rgb}{0.0, 0.0, 0.61}
\definecolor{internationalkleinblue}{rgb}{0.0, 0.18, 0.65}
\definecolor{coolblack}{rgb}{0.0, 0.18, 0.39}
\definecolor{darkblue}{rgb}{0.0, 0.0, 0.55}

\usepackage{comment}
\usepackage{stmaryrd} 

\usepackage{geometry} 
\usepackage{amsmath,amssymb}
\usepackage{bm} 

\usepackage{enumerate} 
\usepackage{amsthm}
\usepackage{mathtools} 

\usepackage{filemod} 

\usepackage{varwidth}

\usepackage{pdflscape}
\usepackage{afterpage}

\makeatletter
\newcommand{\leqnos}{\tagsleft@true\let\veqno\@@leqno}
\newcommand{\reqnos}{\tagsleft@false\let\veqno\@@eqno}
\reqnos
\makeatother

\mathtoolsset{showonlyrefs,showmanualtags}

\theoremstyle{plain}
\newtheorem{thm}{Theorem}[section] 
\newtheorem{mainresult}{Main Result}
\newtheorem{lemma}[thm]{Lemma} 
\newtheorem{prop}[thm]{Proposition} 
\newtheorem{cor}[thm]{Corollary} 
\newtheorem{sass}{Standing Assumption}

\theoremstyle{definition}
\newtheorem{defn}[thm]{Definition} 
\newtheorem{mainresultHHJ}{HHJ Main Result}

\theoremstyle{remark}
\newtheorem{rem}[thm]{Remark}

\newtheoremstyle{boldremark}
{\dimexpr\topsep/2\relax} 
{\dimexpr\topsep/2\relax} 
{}          
{}          
{\scshape} 
{.}         
{.5em}      
{}          

\theoremstyle{boldremark}
\numberwithin{equation}{section}

\newtheoremstyle{condition}
{\topsep}
{\topsep}
{\itshape}
{0pt}
{\bfseries}
{.}
{ }
{\thmname{#1}\thmnumber{ #2}\textnormal{\thmnote{ (#3)}}}

\theoremstyle{condition}
\newtheorem{condition}{Condition}

\usepackage{dsfont}
\newcommand{\R}{\mathds{R}}
\newcommand{\N}{\mathds{N}}

\DeclareMathOperator*{\sgn}{sgn}

\newcommand{\e}{\mathrm{e}}
\renewcommand{\theta}{\vartheta}
\renewcommand{\epsilon}{\varepsilon}
\renewcommand{\P}{\mathbb{P}}

\newcommand{\E}{\mathbb{E}}

\newcommand{\BB}{\mathds{B}}
\newcommand{\CC}{\mathds{C}}

\newcommand{\FF}{\mathds{F}}
\newcommand{\II}{\mathds{I}}
\newcommand{\NN}{\mathds{N}}
\newcommand{\OO}{\mathds{O}}
\newcommand{\PP}{\mathds{P}}
\newcommand{\QQ}{\mathds{Q}}
\newcommand{\RR}{\mathds{R}}
\newcommand{\bS}{\mathds{S}}
\newcommand{\TT}{\mathds{T}}
\newcommand{\UU}{\mathds{U}}
\newcommand{\VV}{\mathds{V}}

\newcommand{\cF}{\mathcal{F}}
\newcommand{\cG}{\mathcal{G}}

\newcommand{\sA}{\mathscr{A}}

\newcommand{\sC}{\mathscr{C}}

\newcommand{\sP}{\mathscr{P}}
\newcommand{\sS}{\mathscr{S}}

\newcommand{\diff}{\mathrm{d}}
\newcommand{\dd}{\,\mathrm{d}}

\newcommand{\1}{\mathbf{1}}

\newcommand*{\EX}[2][]{\E^{#1}\left[ #2 \right ]}
\newcommand*{\cEX}[2][]{\E\left[ #2 \,\middle\vert\,\cF_{#1} \right]}

\newcommand*{\as}[1]{#1\text{-a.s.}}

\newcommand*{\ol}[1]{\overline{#1}}

\newcommand*{\cEXt}[2][]{\widetilde{\E}\left[ #2 \,\middle\vert\ \cF_{#1} \right]}

\usepackage{lipsum}

\definecolor{pipurple}{RGB}{128,0,128}
\definecolor{vargreen}{RGB}{0,150,0}
\definecolor{varyellow}{RGB}{180,120,0}

\makeatletter
\g@addto@macro\bfseries{\boldmath}
\makeatother

\usepackage{caption}
\usepackage{subcaption}

\usepackage{IEEEtrantools}

\usepackage{hyperref}

\usepackage[numbers]{natbib}

\begin{document}
	\title{Proper solutions for Epstein--Zin Stochastic Differential Utility}
	
	\author{Martin Herdegen, David Hobson, Joseph Jerome\thanks{All authors: University of Warwick, Department of Statistics, Coventry, CV4 7AL, UK; \{m.herdegen, d.hobson, j.jerome\}@warwick.ac.uk}}
	\date{\today}
	
	\maketitle
	
\begin{abstract}In this article we consider the optimal investment-consumption problem
	for an agent with preferences governed by Epstein--Zin stochastic
	differential utility (EZ-SDU) who invests in a constant-parameter
	Black--Scholes--Merton market over the infinite horizon. The parameter
	combinations that we consider in this paper are such that the risk
	aversion parameter $R$ and the elasticity of
	intertemporal complementarity $S$ satisfy $\theta\coloneqq\frac{1-R}{1-S}>1$.
	In this sense, this paper is complementary to Herdegen, Hobson and
	Jerome [\href{https://arxiv.org/abs/2107.06593}{arXiv:2107.06593}].
	
	The main novelty of the case $\theta>1$ (as opposed to $\theta\in(0,1)$) is that there is an infinite family of utility processes associated to every nonzero consumption stream. To
	deal with this issue, we introduce the economically motivated notion of a
	{\em proper} utility process, where, roughly speaking, a utility process
	is proper if it is nonzero whenever future consumption is nonzero.
	
	We then proceed to show that for a very wide class of consumption streams $C$, there exists a proper utility process $V$ associated to $C$. Furthermore, for a wide class of consumption streams $C$ ,the proper utility process $V$ is unique.
	Finally, we solve the optimal investment-consumption problem in a constant parameter financial market, where we optimise over the right-continuous attainable consumption streams that have a unique proper utility process associated to them.
\end{abstract}



	
\section{Introduction}

In an infinite-horizon, investment-consumption problem the goal is to maximise the discounted expected utility of consumption, where maximisation takes place over investment strategies in a financial market and nonnegative consumption streams (constrained such that the resulting wealth process is nonnegative for all time). Merton~\cite{merton1969lifetime} solved this problem for a constant parameter financial market and constant relative risk aversion (CRRA) utility. In this article we build on the prior work of the authors~\cite{herdegen2021infinite} and analyse the investment-consumption problem for
Epstein--Zin stochastic differential utility (EZ-SDU). EZ-SDU is a generalisation of time-additive CRRA utility that allows a disentanglement of the agent's risk aversion parameter from their temporal variance aversion preferences; see, for example, Duffie and Epstein \cite{duffie1992stochastic},
Schroder and Skiadas \cite{schroder1999optimal} and Herdegen, Hobson and Jerome \cite{herdegen2021infinite}.

For SDU the utility process is given in implicit form as the solution of a backward stochastic differential equation (BSDE), and before we can attempt to optimise over consumption streams, we must first decide how to associate a utility process to a given consumption stream.
The issue is complicated by the fact that since we work over the infinite horizon there is no terminal condition for the BSDE. For some parameter combinations Herdegen et al.~\cite{herdegen2021infinite} show that there exists a unique utility process for every consumption stream (perhaps taking values in $\overline{\R}$). However, in the parameter combinations studied in this paper, uniqueness fails. Thus, to make progress, we must first decide which utility process to associate to a given consumption stream; only then can we attempt to optimise over investment-consumption pairs. 

The goals of the paper are as follows: first, to illustrate non-uniqueness using constant proportional strategies as an exemplar, showing that non-uniqueness is often the norm, and not an aberration; second, to introduce the economically motivated concept of a {\em proper} solution and argue that it gives the `true' utility process to associate to a consumption stream; third, to prove that for a wide class of consumption streams there exists a (unique) proper utility process; and fourth, to show that if we restrict attention to consumption streams for which there exists a unique proper utility process, we can prove a verification argument for the investment-consumption problem and the natural candidate optimal strategy is optimal.

EZ-SDU is parametrised by two coefficients $R$ and $S$ taking values in $(0,1)\cup(1,\infty)$, corresponding to the agent's risk aversion and their elasticity of intertemporal complementarity (the reciprocal of which is the better known as elasticity of intertemporal substitution). The parameter $\theta\coloneqq\frac{1-R}{1-S}$ is critical. First, it is argued in \cite{herdegen2021infinite} that $\theta>0$ is necessary for the Epstein--Zin SDU equation to have a meaningful solution over the infinite horizon. Second, the mathematics of the problem is vastly different depending on whether $\theta \in (0,1)$ or $\theta \in (1,\infty)$. (The boundary case of $\theta = 1$, or equivalently $R=S$, is CRRA utility.)

The parameter combinations leading to $\theta\in(0,1]$ are studied in \cite{herdegen2021infinite} over the infinite horizon and by Kraft, Seifried and Steffensen \cite{kraft2013consumption}, Seiferling and Seifried \cite{seiferling2016epstein}, Kraft and Seifried \cite{kraft2017optimal} and Matoussi and Xing \cite{matoussi2018convex} over the finite time horizon. In all of these papers, the case $\theta>1$ is avoided due to mathematical difficulties. One (highly desirable) property of EZ-SDU when $\theta\in(0,1]$ is that all evaluable consumption streams are uniquely evaluable -- if there exists a solution to the EZ-SDU equation, then it is necessarily unique. This follows from a comparison theorem (see, for example, \cite[Theorem 9.8]{herdegen2021infinite}) which shows that a \textit{subsolution} to the EZ-SDU equation always lies below a \textit{supersolution}; since a solution is both a subsolution and a supersolution, applying the comparison theorem twice to two such solutions yields uniqueness.

This paper deals with the case $\theta>1$. In this case, the requirements of the comparison theorem in \cite{herdegen2021infinite} are not met.
It is tempting to hope that this is a technical issue and that by being smarter it is possible to extend the comparison theorem to the case $\theta>1$, thus resolving issues of uniqueness. However, this is not the case -- the problem with $\theta>1$ is fundamentally different to the problem with $\theta <1$. When $\theta>1$, it is not just that the comparison theorem fails but rather that non-uniqueness is endemic to the problem. (Indeed the only right-continuous consumption stream with a unique utility process is the zero process.) Note that the same issue of non-uniqueness arises in finite time horizon EZ-SDU as well, unless a nonzero bequest function is added at the terminal time. The main goals of this paper are to illuminate \textit{how} non-uniqueness occurs and to provide an economically motivated criterion for selecting the proper solution, before finally solving the investment-consumption problem.

We begin by studying the utility process associated to constant proportional investment-consumption strategies. In this case, an explicit, time-homogeneous utility process is provided in \cite[Proposition 5.4]{herdegen2021infinite} which is valid in the case $\theta>1$.
However, we show in Section~\ref{sec:constant proportional} that for a constant proportional strategy this solution is not the only solution and there exists an infinite family of (equally explicit, but time-inhomogeneous) utility processes.

It is clear that to be able to formulate Merton's optimal investment-consumption problem for EZ-SDU, there must be a rule which assigns a particular utility process to each consumption stream over which we maximise. Various candidates for this assignation rule are plausible. Perhaps the most obvious choice is the \textit{maximal} utility process. The rationale behind this would be that the agent gets to choose which utility process they associate to a given consumption stream and so they naturally choose the best one. However, when $R>1$ the maximal utility process associated to any consumption stream is the zero process, rendering the problem degenerate. An alternative choice might be the ``game theoretic'' or minimax version of the Merton problem, where the agent maximises over the worst utility process associated to each consumption stream. However, when $R<1$ the minimal utility process associated to any consumption stream is the zero process, again rendering the problem degenerate.
Instead, one of the key contributions of the paper is to introduce the notion of a {\em proper} solution.

It will follow from the discussion of constant proportional investment-consumption strategies in Section~\ref{ssec:family of utility processes} below that if $C$ is the consumption stream which arises from a constant proportional strategy (and if $\E[\int_0^\infty C_s^{1-R} ds ] < \infty$), then for each $T \in [0,\infty]$ there exists a utility process associated to $C$ which is nonzero for $t<T$ and zero for $t \geq T$. (In particular, zero is a solution, and corresponds to $T=0$). Economically, this may be interpreted as saying that the amount consumed after time $T$ has no effect on the agent's utility. This may be considered to be undesirable, and motivates the definition of a proper solution:
a solution $V=(V_t)_{t\geq0}$ (of the defining equation for an EZ-SDU utility process, see \eqref{eq:Epstein--Zin SDU} below) is proper if $\E\big[\int_t^\infty C_s^{1-R} \dd s~\big|~\cF_t\big]>0$ implies that $(1-R)V_t>0$..

The notion of a proper solution is based on the economic idea that strictly positive bounded consumption should imply a nonzero utility. We also introduce other notions of a solution of a more mathematical nature, namely the concepts of an extremal solution and a CRRA-order solution.

The main results of the paper for the case $\theta>1$ are as follows (more precise statements follow as Theorems~\ref{thm:proper solution rc}, \ref{thm:O^1-R_+ subset UP}, \ref{thm:three solutions coincide} and \ref{thm:verification} respectively).

\begin{mainresult}\label{mr:A}
	For a very wide class of consumption streams $C$, there exists a proper utility process $V$ associated to $C$.
\end{mainresult}

\begin{mainresult}\label{mr:B}
	For a wide class of consumption streams, the proper utility process is unique.
\end{mainresult}

\begin{mainresult}\label{mr:C}
	For a wide class of consumption streams, the three solution concepts agree.
\end{mainresult}

\begin{mainresult}\label{mr:D}
	In the constant parameter financial model, if we maximise over attainable consumption streams which have an associated unique proper utility process, then the investment-consumption problem is solved by a constant proportional investment-consumption strategy {\rm(}whose parameters may be identified in terms of the parameters of the EZ-SDU and the financial market{\rm\thinspace)}.
\end{mainresult}

These results can be compared with those of Melnyk, Muhle-Karbe and Seifried~\cite{melnyk2020lifetime}, where we restrict the comparison to the case $\theta>1$ -- the subject of this paper. The main focus of \cite{melnyk2020lifetime} is to understand the impact of transaction costs on the investment-consumption problem under EZ-SDU, but the frictionless case is also covered and \cite{melnyk2020lifetime} presents some of the most complete results in the current literature. Melnyk et al. do not prove any existence results and instead choose to optimise over the abstract class of consumption streams for which a solution exists. 
When $\theta>1$, they further restrict to consumption streams $C=(C_t)_{t\geq0}$ whose EZ-SDU utility process $V=(V_t)_{t\geq0}$ satisfies $(1-R)V_t \geq C_t^{1-R}$ for all $t\geq0$ as well as a transversality condition $\lim_{t\to\infty}e^{-\gamma t}\EX{|V_t|}=0$ for a discount factor $\gamma$. It is unclear exactly how large this class is, but there are many consumption streams which we show to have a unique proper solution and which do not lie in this class. In particular, the bound on $V$ and the transversality condition together rule out the candidate optimal strategy for some parameter combinations, forcing Melnyk et al. to impose additional restrictions on the parameter values \cite[Assumption 3.3]{melnyk2020lifetime} . Finally, their approach only works when $R>1$. A more thorough comparison of our results with those in \cite{melnyk2020lifetime} is provided in Section~\ref{ssec:comparison to MMKS}.

The main results of this paper may also be compared with the prior results of the authors for the case $\theta<1$. In that setting, in \cite{herdegen2021infinite} the authors show

\begin{mainresultHHJ}[Herdegen et al., Theorem 10.8] For every consumption stream $C$ there exists a unique utility process $V$ associated to $C$, if we allow the utility process to take values in $[-\infty,\infty]$.
\end{mainresultHHJ}

\begin{mainresultHHJ}[Herdegen et al., Theorem 11.1]	
	In the constant parameter financial model, if we maximise over attainable consumption streams, then the investment-consumption problem is solved by a constant proportional investment-consumption strategy {\rm(}whose parameters may be identified in terms of the parameters of the EZ-SDU and the financial market{\rm\thinspace)}.
\end{mainresultHHJ}

In particular, the results of this paper are less complete than those of \cite{herdegen2021infinite}: we do not show that there exists a utility process for {\em every} consumption stream, but only a wide class; nor when we consider the investment-consumption problem can we maximise over {\em every} consumption stream, again we can only optimise over a wide class.\footnote{Note however, that even in the context of the classical Merton problem for additive utility, many authors restrict attention to a subclass of consumption streams with nice integrability properties, see the paper by Herdegen, Hobson and Jerome~\cite{herdegen2020elementary}
	for a discussion. In this sense the results of \cite{herdegen2021infinite} are unusual in their completeness.} But this lack of completeness must be set against the additional complexity of the problem we consider here --  the non-uniqueness of the utility process is an unavoidable and major issue.

The remainder of the paper is structured as follows. In Section~\ref{sec:constant proportional}, we introduce EZ-SDU and Merton's investment-consumption problem under EZ-SDU. We show that we may associate to each constant proportional investment-consumption strategy a family of EZ-SDU utility processes indexed by $T\in[0,\infty]$, each of which corresponds to ignoring consumption from time $T$ onwards. In Section~\ref{sec:three solutions}, we introduce the notion of a \textit{proper solution}, an \textit{extremal solution}, and a \textit{CRRA-order solution}. The main results are restated precisely in Section~\ref{sec:maintheorems}.

In Section~\ref{sec:comparison}, we introduce subsolutions and supersolutions to the EZ-SDU equation, which, roughly speaking, differ from solutions by replacing the equality in the EZ-SDU equation with an inequality. We then prove a comparison theorem which provides a sufficient criterion under which subsolutions are dominated from above by supersolutions.

Section~\ref{sec:O-solutions} is dedicated to proving the existence of a \textit{CRRA-order solution} via analysis of a transformation of the original problem to one where it is possible to guarantee existence via a contraction mapping and a fixed point argument.
Here, a CRRA-order solution, for a consumption stream $C$, is a utility process $V$ such that there exists constants $k,K \in (0,\infty)$ such that for all $t$,
$k \E\big[\int_t^\infty \frac{C_s^{1-R}}{1-R} \dd s~\big|~\cF_t\big] \leq V_t \leq K \E\big[\int_t^\infty \frac{C_s^{1-R}}{1-R} \dd s~\big|~\cF_t\big]$. In Section~\ref{sec:extremal solution}, we prove existence and uniqueness of the extremal solution for a class of consumption streams.

Section~\ref{sec:proper solution} considers existence and uniqueness of proper solutions. Furthermore, we show that all three solution concepts agree for consumption streams that are bounded above and below by constant multiples of constant proportional strategies.

In Section~\ref{sec:verification} we verify that the candidate solution proposed in Section~\ref{sec:constant proportional} is optimal over the class of attainable consumption streams to which we may assign a unique proper solution, thus completing the study of the infinite-horizon investment-consumption problem for EZ-SDU.

Appendix \ref{sec:proofs existence of proper} contains some very technical results on the existence of proper solutions.

\section{Epstein--Zin stochastic differential utility and the Merton investment-consumption problem}\label{sec:constant proportional}

Throughout this paper, we work on a filtered probability space $(\Omega, {\mathcal F},  (\cF_t)_{t \geq 0}, \P)$
such that $(\cF_t)_{t\geq0}$ is complete and continuous\footnote{This is slightly stronger than the \textit{right-continuity} assumed in the \textit{usual conditions}. However, it is a necessary assumption for the existence arguments in Appendix \ref{sec:proofs existence of proper} to go through. Whilst we do not believe that left-continuity of the filtration is a necessary assumption for existence of a proper solution to hold in general, it is a pragmatic and convenient assumption, and it does cover the main application of a Brownian filtration and a constant parameter Black-Scholes-Merton financial market.} and $\cF_0$ is $\P$-trivial. Let $\sP$ denote the set of progressively measurable processes, and let $\sP_+$ and $\sP_{++}$ be the restrictions of $\sP$ to processes that take nonnegative and positive values, respectively. Moreover, denote by $\sS$ the set of all semimartingales. We identify processes in $\sP$ or $\sS$ that agree up to indistinguishability.

\subsection{Stochastic differential utility}

To understand Epstein--Zin stochastic differential utility (EZ-SDU), it is beneficial to introduce stochastic differential utility (SDU) in its more general form. We will contrast SDU with \textit{time-additive utility}. We consider infinite-horizon (lifetime) stochastic differential utility.

Time-additive (expected) utility is characterised by a utility function $U:\R_+\times\R_+\to\VV$, where $\VV \subseteq \ol{\R}$ is the domain of the utility function. The utility of a consumption stream $C=(C_t)_{t \geq 0}$ is given by $\EX{\int_0^\infty U(t,C_t)\dd t}$ (provided this expectation is well-defined) and the \textit{utility process} $V=(V_t)_{t \geq 0}$ -- which measures the utility starting from a given time -- is given by $V_t=\cEX[t]{\int_t^\infty U(s,C_s)\dd s}$. Under SDU, the utility function $U$ is generalised to become an \textit{aggregator} $g:\R_+\times\R_+\times\VV\to\VV$. The SDU process $V^C=(V^C_t)_{t\geq0}$ associated to a consumption stream $C$ and an aggregator $g$ is then the solution to
\begin{equation}\label{eq:stochastic differential utility aggregator g}
	V_t = \cEX[t]{\int_t^\infty g(s,C_s,V_s)ds}, \quad t\geq0.
\end{equation}
This creates a feedback effect in which the value at time $t$ may depend in a non-linear way on the value at future times and permits the modelling of a much wider range of preferences. However, in addition to issues about whether the conditional expectation is well-defined, there are new issues concerning the existence and uniqueness of solutions to \eqref{eq:stochastic differential utility aggregator g} which are not present for additive utilities.

\begin{defn}\label{defn:integrable set I(g,C)}
	An \textit{aggregator} is a function $g:[0,\infty)\times \R_+ \times \VV \to \VV$.
	For $C\in\sP_+$, define $\II(g,C)\coloneqq\left\{V\in\sP:~ \E\int_0^\infty \left| g(s,C_s,V_s)\right| \dd s < \infty\right\}$. Further, let $\UU\II(g,C)$ be the set of elements of $\II(g,C)$ which are uniformly integrable. Then $V \in \II(g,C)$ is a \textit{utility process} associated to the pair $(g,C)$ if it has c\`adl\`ag paths and satisfies \eqref{eq:stochastic differential utility aggregator g} for all $t \in [0,\infty)$.
\end{defn}

\begin{rem}\label{rem:utility process UI}
	It can be easily shown that all utility processes are uniformly integrable (see \cite[Remark 3.2]{herdegen2021infinite}).
\end{rem}

\subsection{Epstein--Zin stochastic differential utility}\label{ssec:EZ-SDU}
Epstein--Zin stochastic differential utility (see \cite{duffie1992stochastic,herdegen2021infinite,kraft2014stochastic,schroder1999optimal}) is parameterised by $R,S\in(0,1)\cup(1,\infty)$ and, with $\VV=(1-R)\R_+$, has time-homogeneous aggregator $f_{EZ}:\R_+\times\VV\to\VV$ defined as
\begin{equation}\label{eq:Epstein--Zin aggregator}
	f_{EZ}(c,v) = \frac{c^{1-S}}{{1-S}}\left((1-R)v\right)^\frac{S-R}{1-R}.
\end{equation}
It is convenient to introduce the parameters $\theta=\frac{1-R}{1-S}$ and $\rho=\frac{S-R}{1-R}=\frac{\theta-1}{\theta}$ so that $f_{EZ}(c,v) = \frac{c^{1-S}}{{1-S}}\left((1-R)v\right)^\rho$. Note that when $S=R$ the aggregator reduces to the discounted CRRA utility function $U(c)=\frac{c^{1-R}}{1-R}$. This case corresponds to $\theta=1$ and $\rho=0$.

\begin{rem}
	(a)	In \cite{herdegen2021infinite}, the authors begin by defining the (time-inhomogeneous) Epstein--Zin aggregator $g_{EZ}:\R_+ \times \R_+\times\VV\to\VV$, via
	\begin{equation}\label{eq:Epstein--Zin aggregator previous paper}
		g_{EZ}(t,c,v) = b e^{-\delta t}\frac{c^{1-S}}{{1-S}}\left( (1-R)v\right)^\frac{S-R}{1-R}.
	\end{equation}
	In addition to \eqref{eq:Epstein--Zin aggregator}, this aggregator incorporates a multiplier $b$ and a discount factor $e^{-\delta t}$. However, a simple scaling argument shows that it is possible to set $b=1$ without loss of generality and the authors show in \cite[Section~5.2]{herdegen2021infinite} that the discount factor can be absorbed into the financial market by a change of num\'eraire. Hence, we may also assume $\delta=0$ without loss of generality yielding the aggregator in \eqref{eq:Epstein--Zin aggregator}.
	
	(b) One of the advantages of the aggregator $f_{EZ}$ (or $g_{EZ}$) is that it is one-signed and hence the conditional expectation $\cEX[t]{\int_t^\infty g(s,C_s,V_s)ds}$ is always well defined in $[-\infty,0]$ or $[0,\infty]$. Other authors (e.g. \cite[Appendix C]{duffie1992stochastic} or \cite{melnyk2020lifetime}) have used a slightly different aggregator $g_{EZ}^\Delta$, but in general existence of a utility process for $g^\Delta_{EZ}$ implies existence of a utility process for $f_{EZ}$ and not vice versa. A fuller discussion of this issue can be found in \cite[Section~5.2]{herdegen2021infinite}.
\end{rem}

\begin{defn}\label{def:SDU}
	A process $V^C=V=(V_t)_{t\geq0}$ is an EZ-SDU utility process associated to consumption $C$ if $\int_0^\infty\frac{C_s^{1-S}}{{1-S}}\left( (1-R)V_s\right)^\rho \dd s \in L^1$ and if, for each $t \geq 0$ it solves
	\begin{equation}\label{eq:Epstein--Zin SDU}
		V_t = \E\left[\left.\int_t^\infty\frac{C_s^{1-S}}{{1-S}}\left( (1-R)V_s\right)^\rho \dd s\right| \cF_t \right].
	\end{equation}
\end{defn}

Implicit in the form of $f_{EZ}$ is the fact that in order to define $((1-R)V)^\rho$ we must have $\sgn(V) = \sgn(1-R)$ and hence that $\VV \subseteq (1-R)\overline{\R}_+$. Then, for there to be equality in \eqref{eq:Epstein--Zin SDU} we must have at least that the signs of the left-hand and right-hand sides of \eqref{eq:Epstein--Zin SDU} agree, i.e. $\sgn(V) = \sgn(1-S)$. This forces $\sgn(1-R)= \sgn(1-S)$ or equivalently $\theta>0$. Some authors have considered the case $\theta<0$, but as argued in \cite[Section~7.3]{herdegen2021infinite}, the solutions they find are akin to bubbles, and are not economically justified. For the rest of the paper we will assume $\theta>0$.

The case $0<\theta<1$ was considered in \cite{herdegen2021infinite} and when $\theta=1$, EZ-SDU reduces to the widely-studied CRRA utility.
In this paper, we consider parameter combinations leading to $\theta>1$; either we have $R<S<1$ or $1<S<R$.
\begin{sass}
	The parameters $R$ and $S$ are such that $\theta\coloneqq\frac{1-R}{1-S}>1$.
\end{sass}

\subsection{The financial market and attainable consumption streams}\label{ssec:financial market}
In this paper, we shall consider a frictionless Black--Scholes--Merton financial market. The market consists of a risk free asset with interest rate $r\in\R$, whose price process $S^0=(S^0_t)_{t\geq0}$ is given by $S^0_t = S^0_0 e^{r t}$, and a risky asset given by geometric Brownian motion with drift $\mu\in\R$, volatility $\sigma>0$ and initial value $S_0 = s_0>0$. Explicitly, the price process $S=(S_t)_{t \geq 0}$ of the risky asset is given by $S_t = s_0\exp(\sigma B_t + (\mu-\frac{1}{2}\sigma^2))$, for a Brownian motion $B=(B_t)_{t\geq0}$.

At each time $t\geq0$, the agent chooses to consume at a rate $C_t\in\R_+$ per unit time, and to invest a proportion $\Pi_t\in\R$ of their wealth into the risky asset. The proportion that they invest in the risk free asset $S^0$ at time $t$ is then given by $1-\Pi_t$. It follows that the wealth process of the agent $X=(X_t)_{t\geq0}$ satisfies the SDE
\begin{equation}\label{eqn:wealth process}
	\dd X_t =  X_t \Pi_t  \sigma \dd B_t + \left( X_t (r + \Pi_t (\mu - r))  - C_t\right)\mathrm{d} t,	
\end{equation}
subject to the initial condition $X_0=x$, where $x>0$ is the agent's initial wealth.
\begin{defn}
	Given $x>0$ an \emph{admissible investment-consumption strategy} is a pair $(\Pi,C)= (\Pi_t,C_t)_{t \geq 0}$ of progressively measurable processes, where $\Pi$ is real-valued and $C$ is nonnegative, such that the SDE \eqref{eqn:wealth process} has a unique strong solution $X^{x, \Pi, C}$ that is $\as{\P}$ nonnegative. We denote the set of admissible investment-consumption strategies for $x > 0$ by $\sA(x)$.
\end{defn}
Since the value associated to a strategy only depends upon consumption, and not upon the amount invested in each of the assets, we introduce the following definition:
\begin{defn}
	A consumption stream $C \in\sP_+$ is called \emph{attainable} for initial wealth $x > 0$ if there exists a progressively measurable process $\Pi = (\Pi_t)_{t\geq0}$ such that $(\Pi, C)$ is an admissible investment-consumption strategy. Denote the set of attainable consumption streams for $x  > 0$ by $\sC(x)$.
\end{defn}

\subsection{The Merton investment-consumption problem for Epstein--Zin stochastic differential utility}
The Merton investment-consumption problem combines elements from Sections \ref{ssec:EZ-SDU} and \ref{ssec:financial market} to consider the problem facing an agent with preferences governed by EZ-SDU who may choose an admissible investment-consumption strategy so as to maximise their subjective utility. Therefore, the goal of the Merton investment-consumption problem is to find
\begin{equation}\label{eq:Control problem first formulation}
	V^*(x)=\sup_{C\in\sC(x)}V^C_0,
\end{equation}
where $V^C=V=(V_t)_{t\geq0}$ solves \eqref{eq:Epstein--Zin SDU}.

However, \eqref{eq:Epstein--Zin SDU} may not possess a unique solution, so the problem in \eqref{eq:Control problem first formulation} is not well formulated until we have decided which utility process to assign to each and every consumption we want to optimise over. Section~\ref{ssec:family of utility processes} illustrates what can go wrong by providing a family of utility processes associated to a simple constant proportional investment-consumption strategy. We will then explain in Section~\ref{sec:three solutions} how to select the economically meaningful utility process from the many available and how this impacts the control problem \eqref{eq:Control problem first formulation}. We reformulate the control problem in Section~\ref{sec:verification}.

\subsection{An explicit utility process associated to constant proportional strategies}
From the scaling properties of the problem it is to be expected that the optimal strategy is to invest a constant proportion of wealth in the risky asset, and to consume a constant proportion of wealth. Consider, therefore, the investment-consumption strategy $\Pi \equiv \pi\in\R$ and $C \equiv \xi X$ for $\xi\in\RR_+$. Fixing a proportional strategy determined by $(\pi,\xi)$, and using It\^o's lemma and the dynamics of $X$ given in \eqref{eqn:wealth process}, we find
\begin{equation}
	X_t^{1-R} = \ x^{1-R} \exp\left(\pi \sigma (1-R) B_t + (1-R)\left({r} + (\mu-r)\pi - \xi - \frac{\pi^2\sigma^2}{2}\right)t\right).
\end{equation}
In particular, $\cEX[t]{X_s^{1-R}} = X_t^{1-R} e^{-H(\pi,\xi)(s-t)}$, where
\begin{equation}
	\label{eq:H(pi,xi)>0}
	H(\pi, \xi) =(R-1)\left({r} + \lambda (\mu-r)\pi - \xi - \frac{\pi^2\sigma^2}{2}R\right).
\end{equation}

Let $\eta = \frac{S-1}{S} \left(r +\frac{\lambda^2}{2R}\right)$ and define $\hat{V}:\R_+\to\R$ by
\begin{equation} \label{eq:hat V candidate}
	\hat{V}(x) = \eta^{-\theta S }\frac{x^{1-R}}{1-R}.
\end{equation}
Then, the following proposition holds.
\begin{prop}[{[}\cite{herdegen2021infinite},~Proposition 5.4{]}]\label{prop:derivecandidate}
	Define $D= \{ (\pi, \xi)\in\R\times\R_{++} : H(\pi, \xi)>0 \}$. Consider constant proportional strategies with parameters $(\pi,\xi)\in D$.        Suppose $\theta>0$ and $\eta  > 0$. Then,
	\begin{enumerate}[{\rm(}i{\rm)}]
		\item For $(\pi,\xi)\in D$, one EZ-SDU utility process $V=(V_t)_{t\geq0}$ associated to the strategy $(\pi,\xi X)$ is given by
		\begin{equation}\label{eq:valfungenstrat2}
			V_t =  h(\pi,\xi)X_t^{1-R}, \qquad \text{where}\qquad h(\pi,\xi) = \frac{\xi^{1-R}}{1-R}  \left( \frac{\theta }{H(\pi,\xi)} \right)^\theta.
		\end{equation}
		\item The global maximum of $h(\pi,\xi)$ over the set $D$ is attained at $(\pi, \xi) = (\frac{\lambda}{\sigma R}, \eta)$ and the maximum is $\frac{\eta^{-\theta S}}{1-R}$.
		\item Suppose we only consider constant proportional strategies, and suppose that to each such strategy we associate the utility process given in \eqref{eq:valfungenstrat2}. Then the optimal strategy is $(\hat{\pi},\hat{\xi}) = (\frac{\lambda}{\sigma R}, \eta)$ and satisfies $V^{\hat{\pi},\hat{\xi}X}_0 = \eta^{-\theta S }\frac{x^{1-R}}{1-R}= \hat{V}(x)$, where $x$ denotes initial wealth.
	\end{enumerate}
\end{prop}
In particular, Proposition \ref{prop:derivecandidate}(i) gives us one explicit utility process associated to the constant strategy $(\pi,\xi X)$ . However, the next section shows that it is far from being the unique utility process. Proposition \ref{prop:derivecandidate}(iii) gives us a candidate optimal strategy.

\begin{rem}\label{rem:constantproportionalsatisfiesselforder}
	For future reference note that, provided $H(\pi,\xi)>0$, if $C$ is the constant proportional consumption stream associated with parameters $(\pi,\xi)$, then $C^{1-R}$ is a geometric Brownian motion and $\cEX[t]{\int_t^\infty C_s^{1-R} ds} = \frac{1}{H(\pi,\xi)}C_t^{1-R}$, for all $t\geq0$.
\end{rem}

\subsection{A family of utility processes indexed by absorption time}\label{ssec:family of utility processes}

In this section we show that when $\theta>1$ for each proportional consumption stream $(\pi,\xi X)$, for $\pi\in\R$, $\xi\in\R_{++}$ such that $H(\pi,\xi)>0$, there exists a family of utility processes, parametrised by the first time they hit zero (and are absorbed).

We postulate a time-dependent form of the utility process $V=(V_t)_{t \geq 0}$ given by $V_t = \frac{A(t)\xi^{1-R}}{1-R}X_t^{1-R}$ for a nonnegative process $A=(A(t))_{t\geq0}$.
Finding a solution associated to the constant proportional investment-consumption strategy with parameters $(\pi,\xi)$ then becomes that of solving
\begin{equation}\label{eq:proportional pi xi integral eq absorption time T}
	\frac{A(t)\xi^{1-R}}{1-R}X_t^{1-R} = V_t
	= \cEX[t]{\int_t^\infty \frac{\xi^{1-S}}{1-S} X_s^{1-S} \left(A(s) \xi^{1-R}X_s^{1-R}\right)^\rho \dd s }	
\end{equation}
Using the expression for $\E[X_s^{1-R}|\cF_t]$, we find that $A=(A(t))_{t\geq0}$ solves the integral equation
$	A(t)e^{-H(\pi,\xi)t} = \theta\int_t^\infty e^{-H(\pi,\xi)s} A(s)^\rho \dd s $
and taking derivatives with respect to $t$ shows that $A$ solves the 
the ODE
\begin{equation}\label{eq:ODE}
	A'(t) = H(\pi,\xi)A(t) - \theta A(t)^\rho.
\end{equation}
Note that one solution is the constant solution $A(t)= \big(\frac{\theta}{H(\pi,\xi)}\big)^\theta$. More generally,
the ODE~\eqref{eq:ODE} is separable and can be solved to give
\begin{equation}\label{eq:Adef}
	A(t)  = \left(  \frac{\theta - \left(\theta - H(\pi,\xi)A(0)^{1/\theta}\right) e^{\frac{H(\pi,\xi)}{\theta}t}}{H(\pi,\xi)}  \right)^\theta
\end{equation}
If we assume that $A(0)<\big(\frac{\theta}{H(\pi,\xi)}\big)^\theta$, then $A$ hits zero at $t = T\coloneqq\frac{\theta}{H(\pi,\xi)} \ln \big( \frac{\theta}{\theta - H(\pi,\xi) A(0)^{1/\theta}} \big)$.
Since $(A(t))_{t\geq T}\equiv0$ is a solution on $[T,\infty)$ we can define a family of solutions to \eqref{eq:ODE} and hence to \eqref{eq:Epstein--Zin SDU}, indexed by $A(0)$ such that
\[ A(t) = \left\{ \begin{array}{lll}  \left(  \frac{\theta - \left(\theta - H(\pi,\xi)A(0)^{1/\theta}\right) e^{\frac{H(\pi,\xi)}{\theta}t}}{H(\pi,\xi)}  \right)^\theta & \; &t < T = \frac{\theta}{H(\pi,\xi)} \ln \left( \frac{\theta}{\theta - H(\pi,\xi) A(0)^{1/\theta}} \right) \\
	0 & & t \geq T \end{array} \right. \]
(Note that if $A(0)> \big(\frac{\theta}{H(\pi,\xi)}\big)^{\theta}$ then $A$ diverges to $\infty$, but this is not consistent with the fact that $\E[V_t] \rightarrow 0$.)
Alternatively, the family of solutions can be thought of as indexed by $T$ where $T = \inf\{t\geq0:V_t=0\}$. Effectively, for the solution indexed by $T$, consumption after $T$ does not yield any utility, and the utility process is zero thereafter.

It is hard to argue, when considering the infinite time horizon, that this represents an economically meaningful reduction of the problem. Hence, intuitively the ``correct'' utility process should correspond to $T=\infty$ and be given by \eqref{eq:valfungenstrat2}.

\begin{rem}
	This issue also arises when trying to evaluate finite-horizon EZ-SDU. A variant of the Merton problem for finite-horizon EZ-SDU and $\theta>1$ is considered by \cite{matoussi2018convex,schroder1999optimal,seiferling2016epstein,xing2017consumption}, among others.
	In \cite{matoussi2018convex,seiferling2016epstein,xing2017consumption} the issue of uniqueness is finessed by incorporating a strictly positive bequest function $U_\epsilon(C_T) = \epsilon\frac{C_T^{1-R}}{1-R}$ with $\epsilon>0$ at the finite time horizon $T$ (and either restricting to consumption streams that are strictly positive or only considering the case $R>1$), so that the EZ-SDU equation in this case becomes 	
	\begin{equation}\label{eq:finite horizon with bequest}
		V_t = \E\left[\left.\int_t^T\frac{C_s^{1-S}}{{1-S}}\left( (1-R)V_s\right)^\rho \dd s + U_\epsilon(C_T)\right| \cF_t \right], \quad \text{for all }0\leq t\leq T.
	\end{equation}
	This is not a viable approach in the infinite-horizon case, as a bequest ``at infinity'' has no meaning. In \cite{schroder1999optimal}, the authors claim that EZ-SDU utility processes are unique by finding a solution to \eqref{eq:finite horizon with bequest}, letting $\epsilon\searrow0$, and then claiming that the limiting process -- which is an EZ-SDU utility process for \eqref{eq:finite horizon with bequest} with zero bequest -- is the unique EZ-SDU utility process. As we have seen in this section, this is not the case.
\end{rem}

The approach taken in this paper is to embrace the multiple utility processes and distinguish the \textit{proper} utility process (which we introduce in the next section) from other utility processes. The aim of the proper utility process is to rule out solutions which
ignore the utility gained from consumption from some finite time onwards.

\section{Three solution concepts and three spaces of consumption streams}\label{sec:three solutions}	

\subsection{Solution concepts}\label{ssec:solutionconcepts}
The following definition of a \textit{proper} solution to the EZ-SDU equation is motivated by the arguments of the previous section.
\begin{defn}\label{def:proper}
	Let $C\in\sP_+$ and suppose that $V=(V_t)_{t\geq0}$ is a solution to \eqref{eq:Epstein--Zin SDU}. Then $V$ is called a \textit{proper solution} if $(1-R)V_t>0$ on $\{\cEX[t]{\int_t^\infty C^{1-R}_s  \dd s } > 0\}$ for all $t\geq 0$ 
	up to null sets.
\end{defn}
The notion of a proper solution immediately excludes the time-inhomogeneous utility processes found in Section~\ref{ssec:family of utility processes}.
\begin{rem}
	Note that $\cEX[t]{\int_t^\infty C^{1-R}_s  \dd s } > 0$ on $\{(1-R)V_t>0\}$ up to null sets for all $t \geq 0$.  Indeed, seeking a contradiction, suppose there are $t \geq 0$ and $D\in\cF_t$ with $\P[D] > 0$ such that $(1-R)V_t>0$ on $D$ but ${\1_D \cEX[t]{\int_t^\infty C_s^{1-R} \dd s} =0}$ $\as{\P}$ Then, $\EX{\1_D\int_t^\infty C_s^{1-S} \dd s}  = 0$ and hence $C=0$ for $\P \otimes \diff t$ almost every $(\omega, t)$ in $D \times [t,\infty)$. Consequently, $\E[\1_D V_t] = \EX{\1_D \int_t^\infty f_{EZ}(C_s,V_s) \dd s} = 0$, and we arrive at a contradiction.
\end{rem}

We also define the \textit{extremal} solution. This corresponds to either the maximal solution or the minimal solution -- depending on the sign of $\VV=(1-R)\R_+$. Such solutions frequently appear in BSDE theory; see e.g. \cite{drapeau2013minimal,drapeau2016dual,peng1999monotonic}.
\begin{defn}\label{def:extremal solution}
	Let $C\in\sP_+$ and suppose that $V=(V_t)_{t\geq0}$ is a solution to \eqref{eq:Epstein--Zin SDU}.
	$V$ is an \textit{extremal} solution if $(1-R)V\geq(1-R)Y$ for any other solution $Y=(Y_t)_{t\geq0}$.
\end{defn}	
\begin{rem}\label{rem:maximal solution proper}
	If there exists a proper solution, then extremal solutions are proper. This follows from Definitions \ref{def:proper} and \ref{def:extremal solution}; if $Y$ is a proper solution and $V$ is an extremal solution, then for each $t \geq 0$, up to null sets,
	\begin{equation}
		(1-R)V_t\geq(1-R)Y_t > 0 \text{ on } \left\{\cEX[t]{\int_t^\infty C^{1-R}_s  \dd s } > 0 \right\}.
	\end{equation}
\end{rem}

The third solution concept that will be useful is a CRRA-order solution.
\begin{defn}\label{def:O equivalence relation}
	Suppose that $X=(X_t)_{t\geq0}$ and $Y=(Y_t)_{t\geq0}$ are nonnegative progressive processes. We say that $X$ has the same order as $Y$ (and write $X \stackrel{\OO}{=} Y$, noting that $\stackrel{\OO}{=}$ is an equivalence relation) if there exist constants $k,K\in(0,\infty)$ such that
	\begin{equation}\label{eq:O(Y) condition}
		0\leq k Y \leq X \leq K Y.
	\end{equation}
	Denote the set of progressive processes with the same order as $Y$ by $\OO(Y)\subseteq\sP_+$. Further, denote by $\OO(Y; k,K)$ the set of processes such that \eqref{eq:O(Y) condition} holds for a pre-specified $k$ and $K$.
\end{defn}

For $X \in \sP_+$, define $J^{X} = (J^{X})_{t\geq0}$ by $J^{X}_t = \cEX[t]{\int_t^{\infty} X_s ds}$.

\begin{defn}
	Let $C\in\sP_+$ and suppose that $V=(V_t)_{t\geq0}$ is a solution to \eqref{eq:Epstein--Zin SDU}. We say that $V$ is a CRRA-order solution if $(1-R)V \stackrel{\OO}{=} J^{C^{1-R}}$.
\end{defn}

\begin{rem}
	If $V$ is a CRRA-order solution, then $V$ is a proper solution.
\end{rem}

\subsection{Spaces of consumption streams}
Define $\BB\CC^{\pi,\xi}$ to be the class of consumption streams that are the same order as a constant proportional strategy with parameters $(\pi,\xi)$,
\(	\BB\CC^{\pi,\xi} \coloneqq \left\{C\in\sP_{++}:~ C \stackrel{\OO}{=} \xi X^{\pi,\xi X} \right\} 
\),
and set $\BB\CC = \cup_{\pi,\xi : H(\pi,\xi)>0} \BB\CC^{\pi,\xi}$. Then, $\BB\CC$ is the set of consumption streams which are the same order as a constant proportional strategy with a finite utility process. Here, $X^{\pi,\xi X}$ is an abbreviation of $X^{1,\pi,\xi X}$ and will be used in cases, where the initial wealth is not important.
By Remark~\ref{rem:constantproportionalsatisfiesselforder}, if $C=\xi X^{\pi,\xi X}$ is a  constant proportional investment-consumption strategy with $H(\pi,\xi)>0$ then $C^{1-R} \stackrel{\OO}{=} J^{C^{1-R}}$.

Let $\bS\OO$ be the class of positive progressively measurable processes of \textit{self-order}, i.e. such that $X$ is of the same order as $J^X$. More precisely, define
\begin{equation}
	\bS\OO \coloneqq \left\{X\in\sP_{++}:~ \E \left[ \int_0^\infty X_t \dd t\right] < \infty \mbox{ and } X \stackrel{\OO}{=} J^X\right\}.
\end{equation}
Furthermore, define $\bS\OO(k,K) = \left\{X\in\sP_{++}:~ \E \left[ \int_0^\infty X_t \dd t\right] < \infty \mbox{ and } k J^X \leq X \leq K J^X\right\}$.
On some occasions we need a slightly stronger condition. Define
$$\bS\OO_\nu  \coloneqq \left\{X\in\bS\OO:~  \big(e^{\nu t} X_t\big)_{t \geq 0} \in \bS\OO \right\}\quad \text{for }\nu\geq0, \qquad \text{and} \qquad \bS\OO_+ = \cup_{\nu > 0} \bS\OO_\nu.$$
For arbitrary $\alpha\in\R$, let $\bS\OO^\alpha$, $\bS\OO^\alpha_\nu$ and $\bS\OO^\alpha_+$ be the sets of processes $X$ such that $X^\alpha$ is in $\bS\OO$, $\bS\OO_\nu$ and $\bS\OO_+$ respectively. In particular, for a pair $(\pi,\xi)$ such that $H(\pi,\xi)>0$ we have that $C = \xi X^{x,\pi,\xi X} \in \bS \OO^{1-R}$.

\begin{rem}\label{rem:YinSO implies KYinSO}
	If $Y\in\bS\OO$, then $K Y\in\bS\OO$ for any $K>0$.
\end{rem}

The following two lemmas provide some set inclusions.
\begin{lemma}\label{lem:SO^mu subset SO^nu}
	For $0 \leq \mu \leq \nu$, $\bS\OO_\mu \supseteq \bS\OO_\nu$.
\end{lemma}
\begin{proof}
	Suppose that $X\in\bS\OO_\nu$ and $\mu < \nu$. Then, there exists $k>0$ such that
	\begin{equation}\label{eq:SO_nu decreasing k}
		X_t \geq k \cEX[t]{\int_t^\infty e^{\nu(s-t)}X_s \dd s}\geq k \cEX[t]{\int_t^\infty e^{\mu(s-t)} X_s\dd s}.
	\end{equation}
	Furthermore, since $X\in\bS\OO$,
	\begin{equation*}
		X_t \leq K \cEX[t]{\int_t^\infty X_s\dd s} \leq K \cEX[t]{\int_t^\infty e^{\mu(s-t)} X_s\dd s}.\qedhere
	\end{equation*}	
\end{proof}
\begin{rem}\label{rem:bounds increasing in nu}
	For each $\nu\geq0$, let $(k(\nu),K(\nu))$ be the tightest interval such that $(e^{\nu t}X_t)_{t\geq0}\!\in\!\bS\OO(k(\nu),K(\nu))$. Then, $k(\nu)$ is decreasing in $\nu$ by \eqref{eq:SO_nu decreasing k}. Furthermore, since $X\in\bS\OO$, $k(\nu)\leq k(0)<\infty$. By a symmetric argument, one can show that $K(\nu)$ is decreasing in $\nu$ as well.
\end{rem}

\begin{lemma}
	$\BB\CC \subseteq \bS\OO^{1-R}_+ \subseteq \bS\OO^{1-R}$.
\end{lemma}
\begin{proof}
	Take $C\in\BB\CC$. Then, there exists $\pi\in\R$ and $\xi\in\R_{++}$ and $C^\dagger=\xi X^{\pi,\xi X}$ such that ${C\stackrel{\OO}{=}C^\dagger}$. Note that $Z^{(\nu)}$ defined by $Z^{(\nu)}_t=e^{\nu t}(C^\dagger_t)^{1-R}$ is a geometric Brownian motion such that $\E\big[Z^{(\nu)}_t\big\vert\cF_s\big]=Z^{(\nu)}_s e^{(\nu-H(\pi,\xi))(s-t)}$ for $s\geq t$. Let $\nu<H(\pi,\xi)$ and $Z= Z^{(\nu)}$. Then, $Z\stackrel{\OO}{=}J^{Z}$ by Remark \ref{rem:constantproportionalsatisfiesselforder}.
	
	Define $Y$ by $Y_t=e^{\nu t}C^{1-R}$. Then, $Y\stackrel{\OO}{=}Z$ since $C\stackrel{\OO}{=}C^\dagger$. By integrating from time $T$ onwards and taking conditional expectations $J^Y\stackrel{\OO}{=}J^Z$. Since $\stackrel{\OO}{=}$ is an equivalence relation, combining these gives $Y\stackrel{\OO}{=}Z\stackrel{\OO}{=}J^Z\stackrel{\OO}{=}J^Y$ and $Y\in\bS\OO$. Thus, $C\in\bS\OO_\nu^{1-R}\subseteq\bS\OO_+^{1-R}$.
\end{proof}

\section{Main theorems}\label{sec:maintheorems}

The four theorems in this section correspond to the four main results presented in the introduction. Proofs of these and the associated propositions, are given in later sections.

The first proposition states that all consumption streams $C\in\sP_+$ with $C^{1-R}\in\bS\OO$ have a CRRA-order solution associated to them. It is proven in \cite[Theorem 8.6]{herdegen2021infinite} and explicit bounds on $J^{C^{1-R}}((1-R)V)^{-1}$ are derived in Section \ref{sec:O-solutions}.
\begin{prop}\label{prop:O solution for C^1-R in SO}
	Suppose that $C^{1-R}\in\bS\OO$. Then, there exists a unique CRRA-order solution to \eqref{eq:Epstein--Zin SDU}.
\end{prop}
The next result says that we may relax the lower bound on the consumption streams in Proposition~\ref{prop:O solution for C^1-R in SO} and still find a solution $V$. In particular, we may evaluate consumption streams such that $C^{1-R}$ is bounded above by a process in $\bS\OO$. We find an extremal solution in these cases. 

\begin{prop}\label{prop:extremal solution for C^1-R < hatC^1-R}
	For each $C\in\sP_+$ such that $C^{1-R}\leq Y\in\bS\OO_+$, there exists a unique extremal solution $V^C$ to \eqref{eq:Epstein--Zin SDU}. Furthermore, $V^C$ is increasing in $C$.
\end{prop}

We now turn to \textit{proper solutions}. The following result shows that we may find a proper solution associated to a large class of consumption streams.
\begin{thm}\label{thm:proper solution rc}
	Suppose that $C\in\sP_+$ is a right-continuous consumption stream such that ${C^{1-R}\leq Y\in\bS\OO_+}$. Then, there exists a proper solution to \eqref{eq:Epstein--Zin SDU}.
\end{thm}
Proper solutions are an economically meaningful concept that allow us to choose from the many solutions to the EZ-SDU equation and Theorem \ref{thm:proper solution rc} provides a large class of consumption streams which have proper solutions. However, we have not yet discussed their uniqueness. If the property of being proper does not provide a criteria for selecting a unique solution, then it does not help to overcome the issues of non-uniqueness intrinsic to EZ-SDU when $\theta>1$. The following definition is therefore of great importance.
\begin{defn}
	We say that $C\in\sP_+$ is \textit{uniquely proper} if there exists a unique proper solution to \eqref{eq:Epstein--Zin SDU}.
	Let $\UU\PP$ denote the set of uniquely proper consumption streams.
\end{defn}
\begin{thm}\label{thm:O^1-R_+ subset UP}
	$\bS\OO^{1-R}_+ \subseteq\UU\PP$.
\end{thm}
For consumption streams in $\bS\OO^+$, all three solution concepts coincide.
\begin{thm}\label{thm:three solutions coincide}
	Suppose that $C\in\bS\OO^{1-R}_+$. Then, the following three solutions to \eqref{eq:Epstein--Zin SDU} all coincide:
	\begin{enumerate}
		\item the CRRA-order solution associated to $C$ given in Proposition \ref{prop:O solution for C^1-R in SO};
		\item the unique extremal solution;
		\item the unique proper solution.
	\end{enumerate}
\end{thm}
Finally, in Section~\ref{sec:verification}, we prove a verification theorem that shows that the candidate optimal proportional investment-consumption strategy given in Proposition \ref{prop:derivecandidate} is optimal over all attainable and uniquely proper right-continuous consumption streams.
\begin{defn}Let $\UU\PP^*$ be the restriction of $\UU\PP$ to the right-continuous processes.
\end{defn}
\begin{thm}[Verification Theorem]\label{thm:verification}
	Let initial wealth be $x>0$. Suppose that $\theta> 1$ and $\eta>0$. If $V^C$ is the unique proper utility process associated to $C\in\UU\PP^*$ and $\hat{V}(x)$ is the candidate optimal utility given in \eqref{eq:hat V candidate}, then
	\begin{equation}
		\sup_{C \in \sC(x)\cap\UU\PP^*} V^C_0 = V^{\hat{C}}_0 = \hat{V}(x)
	\end{equation}
	and the optimal investment-consumption strategy is given by $(\hat{\Pi},\hat{C}) = (\hat{\pi},\hat{\xi} X^{x,\hat{\pi}, \hat{\xi}X})$ where $\hat{\pi}\coloneqq \frac{\mu-r}{\sigma^2 R}$ and $\hat{\xi}\coloneqq\eta = \frac{S-1}{S} \big(r +\frac{\lambda^2}{2R}\big)$.
\end{thm}

It is interesting to contrast the results of this paper with those of Herdegen et al.~\cite{herdegen2021infinite} and Melnyk et al.~\cite{melnyk2020lifetime}.

\subsection{A comparison with Herdegen et al}

The goals of \cite{herdegen2021infinite} are two-fold; first to discuss how best to formulate the infinite-horizon investment-consumption problem, and second to solve the problem when $0<\theta<1$. The discussion in Part~I of \cite{herdegen2021infinite} motivates the set-up of the problem we use here, including the form of the EZ-aggregator $f_{EZ}$.

In Part~II of \cite{herdegen2021infinite}, the authors consider existence and uniqueness of the utility process and the investment-consumption problem in the case $0<\theta<1$.
The fundamental existence result \cite[Theorem 10.8]{herdegen2021infinite} is reused in this paper to give existence for $\theta>1$ when the consumption stream satisfies a self-order condition. Thereafter, the methodology and approach of \cite{herdegen2021infinite} and this paper differ considerably. In \cite{herdegen2021infinite}, the authors  prove a comparison lemma (\cite[Theorem 9.8]{herdegen2021infinite}) whose use is threefold. First, it gives a simple proof of uniqueness of the solution to the equation for the utility process. Second, it can be used to prove a monotonicity result which can be used to generalise the existence result from the consumption streams satisfying the self-order condition to all consumption streams. Third, the comparison theorem plays a crucial role in the verification lemma.

When $\theta>1$, the hypotheses of the comparison lemma in \cite{herdegen2021infinite} are not satisfied. Indeed, the result cannot be true when $\theta>1$ as then uniqueness would follow -- and we have seen that uniqueness fails in general. Hence, we are required to introduce a new concept to identify the economically relevant solution. We call these proper solutions. Our goal is then to prove existence and uniqueness of such solutions. This requires a new and fundamentally different comparison lemma and new ideas to prove existence and uniqueness.

To prove existence of a proper solution, we first consider the case of a filtration which is constant (with respect to time) except at a finite set of time-points. In that setting, and for a class of particularly simple consumption streams, we give an explicit bound on the utility process associated to a proper solution. This bound can then be extended to a continuous filtration and a wide class of consumption streams, thus showing that there exists a proper solution in this setting.

Uniqueness of a proper solution also needs a completely new argument, and involves showing that given a pair of proper solutions, the ratio of the pair must either stay constant at unity or explode. The verification lemma now follows in a similar spirit to \cite{herdegen2021infinite} and relies on a perturbation argument first given in \cite{herdegen2020elementary}. However, some of the arguments need to be modified since we are discussing proper solutions. In particular, we need to prove an important result which says that we may approximate $C\in\UU\PP$ by a sequence $(C^n)_{n\in\N}$ of consumption streams with associated proper solutions.

\subsection{A comparison with Melnyk et al}\label{ssec:comparison to MMKS}

It is also interesting to compare the results and techniques of this paper with those of the recent and wide-ranging paper by Melnyk et al.~\cite{melnyk2020lifetime}. The ultimate focus of \cite{melnyk2020lifetime} is to consider the investment-consumption problem for EZ-SDU with transaction costs. However, they first consider the frictionless case. They use a slightly different set-up, and a reduced set of parameter combinations -- in particular, they require that $R>1$.

The main result in \cite{melnyk2020lifetime} regarding the optimal investment-consumption strategy (Theorem 3.4) involves calculating the optimal strategy in a space $\sA^0$ \cite[Definition 3.1]{melnyk2020lifetime}. For a strategy to belong to this class, the wealth and investment process must satisfy some integrability conditions, and there must exist a unique utility process. (Although uniqueness of the utility process is considered via a comparison lemma, which is different in both statement and proof to those in this paper and \cite{herdegen2021infinite}, existence is not considered in \cite{melnyk2020lifetime}). Finally, for the case $\theta>1$ results are proved under three extra conditions: the first is that consumption streams $C=(C_t)_{t\geq0}$ must be strictly positive; the second \cite[Equation 5]{melnyk2020lifetime} is that the associated utility process $V=(V_t)_{t\geq0}$ satisfies $(1-R)V_t \geq C_t^{1-R}$ for $t\geq0$; the third \cite[Equation 3]{melnyk2020lifetime} is that the utility process satisfies a transversality condition $\lim_{t\to\infty}e^{- \gamma t}\EX{|V_t|}=0$ for an appropriate $\gamma \in \R$. The second and third of these assumptions are rather ad-hoc conditions but form a crucial element of the proof of the comparison theorem. Combining the first and second of the trio of assumptions ensures that $(1-R)V_t>0$ for all $t\geq0$. This has the effect of identifying the proper solution within the class of utility processes; however, these extra assumptions rule out many consumption streams for which there exists a proper utility process. Furthermore, for certain parameter combinations the second and third assumptions even rule out the candidate optimal strategy. As a consequence, Melynk et al. are forced to require additional restrictions on the parameters \cite[Assumption 3]{melnyk2020lifetime} beyond those which are necessary to ensure that the problem is well-posed.

\section{Subsolutions and supersolutions}\label{sec:comparison}

This section introduces subsolutions and supersolutions and then proves that, under certain conditions, subsolutions are bounded above by supersolutions. This important result will be used in all of the remaining sections.

We recall from \cite{herdegen2021infinite} the definition of an a \textit{aggregator random field} (a generalisation of the aggregator that is allowed to depend on the state of the world $\omega\in\Omega$) as well as the definition of a sub- and supersolution.
\begin{defn}[[\citealp{herdegen2021infinite},~Definition 9.1{]}]
	An \textit{aggregator random-field} $g:[0,\infty) \times \Omega \times \R_+ \times \VV \to \VV$ is a product measurable mapping such that
	$g(\cdot,\omega,\cdot,\cdot)$ is an aggregator for fixed $\omega\in\Omega$, and for progressively-measurable processes $C=(C_t)_{t\geq0}$ and $V=(V_t)_{t\geq0}$, the process $(g(t,\omega,C_t(\omega),V_t(\omega)))_{t\geq0}$ is progressively-measurable.
\end{defn}
We need the more general notion of an aggregator random fields since it permits us to stochastically perturb our Epstein--Zin aggregator.
\begin{defn}[[\citealp{herdegen2021infinite},~Definition 9.3{]}]
	Let $C \in \sP_+$ and $g$ be an aggregator random field such that $g$ takes only one sign, i.e. $\VV\subseteq \ol \R_+$ or $\VV\subseteq \ol \R_-$. A $\VV$-valued l\`ad optional process $V$ is called
	\begin{itemize}
		\item  a \textit{subsolution} for the pair $(g,C)$  if $\limsup_{t\to\infty}~ \EX{V_{t+}} \leq 0$ and for all bounded stopping times $\sigma\leq\tau$,
		\begin{equation}\label{eq:subsolution equation}
			V_\sigma ~\leq~ \cEX[\sigma]{V_{\tau{+}} +  \int_\sigma^{\tau} g(s,\omega, C_s,V_s)\dd s}.
		\end{equation}
		\item   a \textit{supersolution} for the pair $(g,C)$  if $\liminf_{t\to\infty}~ \EX{V_{t+}} \geq 0$ and for all bounded stopping times $\sigma\leq\tau$,
		\begin{equation}\label{eq:supersolution equation}
			V_\sigma ~\geq~ \cEX[\sigma]{V_{\tau{+}} +  \int_\sigma^{\tau} g(s,\omega, C_s,V_s)\dd s}.
		\end{equation}
		\item a \emph{solution} for the pair $(g,C)$ if it is both a subsolution and a supersolution and ${V\in\II(g,C)}$.
	\end{itemize}
\end{defn}
\begin{rem}\label{rem:solutions are utility processes} By taking the limit as $\tau\to\infty$ and using the transversality condition, it is clear that a solution $V$ for the pair $(g,C)$ satisfies \eqref{eq:stochastic differential utility aggregator g}. We then choose a c\`adl\`ag version of $V$ so that $V$ is a utility process associated to $(g,C)$. This implies in particular that if $V$ is a solution for the pair $(g,C)$, then $V_\sigma = \cEX[\sigma]{\int_\sigma^\tau g(s,\omega, C_s,V_s)\dd s +V_{\tau}}$ for all bounded stopping times $\sigma\leq\tau$ since $V_{\tau} = V_{\tau+}$.
\end{rem}

\subsection{Comparison of subsolutions and supersolutions}
It is shown in \cite[Theorem 9.8]{herdegen2021infinite} that when $g$ is decreasing in its last argument, if $V^1$ is a subsolution and $V^2$ is a supersolution (both associated to $g$ and some $C\in\sP_+$) and either $V^1$ or $V^2$ is in $\UU\II(g,C)$, then $V^1_\tau\leq V^2_\tau$ for all finite stopping times $\tau$. However, when $\theta>1$, the Epstein--Zin aggregator $f_{EZ}$ is increasing in its last argument so this theorem does not apply.

The next proposition shows that we may weaken the condition that the aggregator has a negative derivative with respect to its last argument, and instead assume that it has a derivative which is bounded above by some positive decreasing exponential. 

We introduce the following condition on a pair $(V^1,V^2)$ of stochastic processes which will be a requirement for our comparison theorems.
\begin{condition}\label{cond:V^1,V^2 condition}
	Let $g:[0,\infty) \times \Omega \times \R_+ \times \VV \to \VV$ be an aggregator random field and $C\in\sP_+$. The pair $(V^1,V^2)$ satisfies Condition~\ref{cond:V^1,V^2 condition} for the pair $(g,C)$ if either $V^1$ or $V^2$ is in $\UU\II(g,C)$ and $(V^1-V^2)^+$ is $L^1$ bounded.
\end{condition}
\begin{rem}\label{rem:UI stronger than condition A}
	Note that for a subsolution $V^1$ and a supersolution $V^2$ associated to the pair $(g,C)$, a sufficient condition for the pair $(V^1,V^2)$ to satisfy Condition \ref{cond:V^1,V^2 condition} for the pair $(g,C)$ is that $V^1,V^2\in\UU\II(g,C)$. This is because $\EX{(V^1-V^2)^+} \leq |V^1_0|+|V^2_0| + \EX{\int_0^\infty |g(s,\omega, C_s,V^1_s)|\dd s}+\EX{\int_0^\infty |g(s,\omega, C_s,V^2_s)|\dd s} < \infty$. However, Condition \ref{cond:V^1,V^2 condition} is more general.
\end{rem}

\begin{prop}[Comparison Theorem]\label{prop:comparison}
	Let $g:[0,\infty) \times \Omega \times \R_+ \times \VV \to \VV$ be an aggregator random field that is concave and nondecreasing in its last argument. Let $C\in\sP_+$. Suppose that $V^1$ is a subsolution and $V^2$ is a nonzero supersolution for the pair $(g,C)$ and that the pair $(V^1,V^2)$ satisfies Condition~\ref{cond:V^1,V^2 condition} for the pair $(g,C)$. Suppose further that for $\P$-a.e.~$\omega$, $g_v(t,\omega,C_t(\omega),V^2_t(\omega)) \leq \kappa e^{-\nu t}$  for all $t \geq 0$ for some $\kappa,\nu>0$. Here, we interpret $g_v$ to be the right-derivative of $g$ with respect to $v$, which exists everywhere for $v \neq  0$ by the concavity assumption. Then, $V^1_\sigma  \leq V^2_\sigma$ $\as{\P}$ for all finite stopping times $\sigma$.
\end{prop}

\begin{proof}
	Seeking a contradiction, suppose there is a finite stopping time $\sigma$ such that ${\P[V^1_{\sigma} > V^2_{\sigma}] > 0}$. By replacing $\sigma$ with $\sigma \wedge T$ for $T$ sufficiently large, we may assume without loss of generality that $\sigma$ is bounded. Set $A := \{V^1_\sigma > V^2_{\sigma} \}$. Since $V^1$ and $V^2$ are l\`ad, we may define the right-continuous processes $(V^1_{t+})_{t \geq 0}$ and $(V^2_{t+})_{t \geq 0}$. Further, define the stopping time ${\tau} := \inf\{t \geq \sigma:  V^1_{t+} \leq V^2_{t+}\}$. Then $(V^1_{\tau+} - V^2_{\tau+})\1_{\{\tau < \infty\}}\leq 0$ by the right-continuity of $(V^1_{t+})_{t \geq 0}$ and $(V^2_{t+})_{t \geq 0}$.
	
	First, we show that $\P[A \cap \{\sigma < \tau\}] > 0$. Indeed, otherwise if $\1_{\{\sigma = \tau\} \cap A} = \1_A$ $\as{\P}$, the definition of sub- and supersolutions yields
	\begin{eqnarray}
		\label{eq:pf:prop:comparision}\qquad~~
		\1_A  \left(V^1_\sigma - V^2_\sigma \right) \leq \cEX[\sigma]{\1_A \left(V^1_{\sigma+}- V^2_{\sigma+}  \right)} = \cEX[\sigma]{\1_A \left(V^1_{\tau+}- V^2_{\tau+}  \right)\1_{\{\tau < \infty\}} } \leq 0,
	\end{eqnarray}
	and we arrive at a contradiction.
	
	Next, by the definition of sub- and supersolutions and by Jensen's inequality for the convex function $f(x)=x^+ = \max\{x,0\}$, for $t\leq T$, letting $B_t = A\cap\{\sigma\leq t<\tau\}$
	\begin{align}
		\lefteqn{\1_{B_t}\left(V^1_t - V^2_t \right)^+}
		\\
		&\leq \cEX[t]{\1_{B_t}\!\left(V^1_{(T\wedge\tau)+}- V^2_{(T\wedge\tau)+} + \int_t^{T\wedge\tau}\left[g(s,\omega,C_s,V^1_{s})-g(s,\omega,C_s,V^2_{s})\right]\dd s \right)^+}
		\\
		&\leq\cEX[t]{\1_{B_t}\!\big(V^1_{(T\wedge\tau)+}- V^2_{(T\wedge\tau)+}\big)^+\! + \1_{B_t}\!\!\int_t^{T\wedge\tau}\!\!\!\left(g(s,\omega,C_s,V^1_{s})-g(s,\omega,C_s,V^2_{s})\right)^+\!\!\dd s}
	\end{align}
	where the right hand side is well-defined since either $V^1$ or $V^2$ is in $\UU\II(g,C)$.	Taking expectations yields
	\begin{align}
		\lefteqn{\EX{\1_{B_t}\left(V^1_t - V^2_t \right)^+}}
		\\ \label{eq:proof:comparison:1}
		~~~~&~~\leq\EX{\1_{B_t}\!\big(V^1_{(T\wedge\tau)+}- V^2_{(T\wedge\tau)+}\big)^+\! + \1_{B_t}\!\!\int_t^{T\wedge\tau}\!\!\!\left(g(s,\omega,C_s,V^1_{s})-g(s,\omega,C_s,V^2_{s})\right)^+\!\!\dd s}
	\end{align}
	
	Taking the $\limsup$ as $T \to \infty$ and using the fact that $\1_{B_t} \1_{\tau\leq T}\left(V^1_{\tau+}- V^2_{\tau+}\right)^+=0~\as{\P}$ for all $T\geq0$ and the transversality condition of sub- and supersolutions gives
	\begin{eqnarray}
		\lefteqn{\limsup_{T\to\infty}\EX{\1_{B_t}\left(V^1_{(T\wedge\tau)+}- V^2_{(T\wedge\tau)+}\right)^+}}
		\\
		&=&~ \limsup_{T\to\infty}\EX{\1_{B_t}\1_{T<\tau}\left(V^1_{T+}- V^2_{T+}\right)^+} + \limsup_{T\to\infty}\EX{\1_{B_t} \1_{\tau\leq T}\left(V^1_{\tau+}- V^2_{\tau+}\right)^+}
		\\
		&\leq&~ \limsup_{T\to\infty}\EX{\left(V^1_{T+}- V^2_{T+}\right)^+} 
		~\leq~ \limsup_{T\to\infty}\EX{(V^1_{T+})^+ +(V^2_{T+})^-} ~\leq~ 0,
	\end{eqnarray}
	where the last inequality follows since either $\VV=\R_+$ (and $(V^2_{T+})^-=0$) along with the transversality condition for supersolutions, or $\VV=\R_-$ (and $(V^1_{T+})^+=0$) along with the transversality condition for supersolutions.
	Hence, by taking the $\limsup$ as $T\to\infty$ and using the positivity of the integrand, \eqref{eq:proof:comparison:1} becomes
	\begin{eqnarray*}
		\lefteqn{\EX{\1_{B_t}\left(V^1_t - V^2_t \right)^+}}
		\\
		&\leq&~\EX{\1_{B_t}\int_t^\tau\left(g(s,\omega,C_s,V^1_{s})-g(s,\omega,C_s,V^2_{s})\right)^+\dd s},
		\\
		&\leq&\EX{\int_t^\infty \1_{B_s}g_v(s,\omega,C_s,V^2_s)\left(V^1_s - V^2_s \right)^+ \dd s}
		\\
		&\leq&~\EX{\int_t^\infty \kappa e^{-\nu s} \1_{B_s} \left(V^1_s - V^2_s \right)^+ \dd s},
	\end{eqnarray*}
	where in the middle line we have used that $g$ is concave and nondecreasing in its last argument and $V^2_s \neq 0$.
	If $\Gamma(t) \coloneqq\EX{\1_{B_t}\left(V^1_t - V^2_t \right)^+}$, then $\Gamma = (\Gamma(t))_{t\geq0}$ is a nonnegative process such that $\Gamma(t) \leq \int_t^\infty \kappa e^{-\nu s} \Gamma(s) \dd s$.	Note that ${\Gamma(t)\leq \EX{(V^1_t-V^2_t)^+}\leq \gamma}$ for some $\gamma>0$, by the $L^1$-boundedness of $(V^1-V^2)^+$.  Therefore, since $\int_0^\infty \kappa e^{-\nu t}\dd t = \frac{\kappa}{\nu}$ and 
	$\int_0^\infty \kappa e^{-\nu s} \Gamma(s) \dd s\leq \frac{\gamma\kappa}{\nu}$, we can apply Gr\"onwall's inequality for Borel functions (\cite[Theorem 2.5]{herdegen2017minimal} with $y(t)=\Gamma(-t)$ and $\mu(A)=\int_{A\cap\R_-} \kappa e^{\nu t}\dd t$) to conclude that $\Gamma(t)=0$ for all $t > 0$.
	
	Note that $\1_{B_t}\left(V^1_t - V^2_t\right)^+ \geq 0$ for each $t \geq 0$.  Hence, by Fatou's Lemma,
	\begin{equation}\label{eq:pf:prop:comparison3}
		0\leq\EX{\1_{B_t}\left(V^1_{t+} - V^2_{t+} \right)^+} \leq \liminf_{s\downdownarrows t}\EX{\1_{B_s}\left(V^1_s - V^2_s \right)^+}=0.
	\end{equation}	
	Furthermore, since $\1_{B_t}\left(V^1_{t+} - V^2_{t+}\right) = \1_{B_t}\left(V^1_{t+} - V^2_{t+}\right)^+ \geq 0$ for each $t \geq 0$ by the definition of $\tau$, it follows from \eqref{eq:pf:prop:comparison3} that $P_t = \1_{B_t}\left(V^1_{t+} - V^2_{t+}\right) =0$ $\as{\P}$ for all $t\geq0$.
	Since $(V^1_{t+})_{t \geq 0}$, $(V^2_{t+})_{t \geq 0}$ and $\1_{B_t}$ are right-continuous, $P=(P_t)_{t\geq0}$ is right continuous and is therefore indistinguishable from zero. In particular, $\1_A\1_{\sigma < \tau}\left(V^1_{\sigma+} - V^2_{\sigma+} \right) = \1_{B_\sigma}\left(V^1_{\sigma+} - V^2_{\sigma+} \right) = 0$ $\as{\P}$ But then the definition of  sub-and supersolutions implies that
	\begin{eqnarray}
		\1_A\1_{\sigma < \tau}  \left(V^1_\sigma - V^2_\sigma \right) \leq \cEX[\sigma]{\1_A \1_{\sigma < \tau}\left(V^1_{\sigma+}- V^2_{\sigma+}  \right)}  =0,
	\end{eqnarray}
	and we arrive at a contradiction.
\end{proof}
The following corollary will be used in the Verification Theorem later.
\begin{cor}\label{cor:comparison C^1-R < K (1-R)V}
	Let $f_{EZ}$ be the Epstein--Zin aggregator and suppose that $R<1$. Let $C\in\sP_+$. Suppose that $V^1$ is a subsolution and $V^2$ is a positive supersolution for the pair $(f_{EZ},C)$ and that the pair $(V^1,V^2)$ satisfies Condition~\ref{cond:V^1,V^2 condition} for the pair $(f_{EZ},C)$. Suppose further that $C_t^{1-R}\leq K e^{-\gamma t}(1-R)V^2_t$ for some $K,\gamma>0$ and all $t\geq0$. Then, $V^1_\sigma \leq V^2_\sigma$ for all finite stopping times $\sigma\geq0$.
\end{cor}
\begin{proof}
	Taking derivatives of $f_{EZ}$ with respect to its second argument gives for $v > 0$
	\begin{align}
		\label{eq:f_EZ first derivative}
		\dfrac{\partial f_{EZ}}{\partial v}(c, v) =&~ (\theta-1)c^{1-S}((1-R)v)^{-\frac{1}{\theta}} \geq 0,
		\\
		\dfrac{\partial^2 f_{EZ}}{\partial v^2}(c, v) =&~ -\rho (1-R)c^{1-S}((1-R)v)^{-(1+\frac{1}{\theta})} \leq 0.
	\end{align}
	Hence, using \eqref{eq:f_EZ first derivative}, $\frac{\partial f_{EZ}}{\partial v}(C_s, V^2_s) \leq (\theta-1)K^{\frac{1}{\theta}} e^{-\frac{\gamma}{\theta}t}$ and the conditions of Proposition \ref{prop:comparison} are met with $\kappa=(\theta-1)K^\frac{1}{\theta}$ and $\nu=\frac{\gamma}{\theta}$.
\end{proof}
\begin{rem}
	Note that the utility process associated to constant proportional strategies is given in \eqref{eq:valfungenstrat2} by $h(\pi,\xi) X_t^{1-R}$. Since $C_t = \xi X$, the conditions of Corollary \ref{cor:comparison C^1-R < K (1-R)V} are not met, and we cannot use it to give a uniqueness result (even over constant proportional strategies). We instead use Corollary \ref{cor:comparison C^1-R < K (1-R)V} in the proof of the Verification Theorem (Theorem \ref{thm:verification}), in which we perturb the candidate value function beforehand.
\end{rem}

\section{The CRRA-order solution}\label{sec:O-solutions}

This section summarises \cite[Theorem B.2]{herdegen2021infinite} which gives conditions under which a unique CRRA-order solution can be shown to exist. Under such conditions, explicit bounds on the associated utility process are provided. We begin by introducing a simplifying change of coordinates.

Define the nonnegative processes $W = (1-R)V$ and $U = U(C) = \theta C^{1-S}$ and the aggregator
\begin{equation}\label{eq:aggregator h_EZ}
	h_{EZ}(u,w) = u w^\rho.
\end{equation}
Further define $J = J^{U^\theta}$ by
\begin{equation}\label{eq:J^Utheta}
	J_t = \cEX[t]{\int_t^\infty U_s^\theta \dd s}, \quad \text{for all }t\geq0.
\end{equation}
Note that $V \in \II(f_{EZ},C)$ if and only if $ W \in \II(h_{EZ}, U(C))$. Hence, $V^C$ is a utility process associated to consumption stream $C$ with aggregator $f_{EZ}$ if and only if $W=W^{U(C)}$ is a utility process associated to consumption stream $U(C)$ with aggregator $h_{EZ}$.	Furthermore, $V$ is a CRRA-order solution if and only if $W$ is solution such that $W\stackrel{\OO}{=}J$ and $V$ is an extremal solution if and only if $W$ is a \textit{maximal} solution. Finally, $V=(V_t)_{t\geq0}$ is a proper solution associated to $(f_{EZ},C)$ if and only if $W=(W_t)_{t\geq0}$ is a solution associated to $(h_{EZ},U)$ such that $W_t >0 $ on $\{ \cEX[t]{\int_t^\infty U_s^\theta \dd s} >0 \} $  up to null sets for all $t\geq0$. In a slight abuse of the definition, we then also refer to $W$ as being proper.

We will work in this coordinate system until Section~\ref{sec:verification} and prove existence and uniqueness results in these coordinates; translation to the original coordinate system is immediate.

Define the operator $F_U:\II(h_{EZ},U)\to \sP_+$ by
\begin{equation}\label{eq:fixed point operator F}
	F_{U}(W)_t = \cEX[t]{\int_t^\infty U_s W_s^\rho \dd s},\quad \text{for all }t\geq0;
\end{equation}
we always choose a c\`adl\`ag version for the right-hand side of \eqref{eq:fixed point operator F}. Note that $W$ is a solution associated to $(h_{EZ},U)$ if and only if it is a fixed point of the operator $F_U$. To show existence of a \textit{maximal} solution associated to $(h_{EZ},U)$ in Section~\ref{sec:extremal solution}, it will instead be beneficial to prove existence of a fixed point to a more general (perturbed) operator.
\begin{prop}[[\citealp{herdegen2021infinite},~Theorem B.2{]}]\label{prop:existence of a solution, bounded and perturbed}  Let $\epsilon\geq0$ and let $U\stackrel{\OO}{=}\Lambda$ for $\Lambda\in\bS\OO^\theta$. Let $J=(J_t)_{t\geq0}$ be defined by \eqref{eq:J^Utheta}. Then, $F^{\epsilon}_{U,\Lambda}:\II(h_{EZ},U)\to \sP_+$, defined by
	\begin{equation}\label{eq:perturbed fixed point operator}
		F^\epsilon_{U,\Lambda}(W)_t = \cEX[t]{\int_t^\infty U_s W_s^\rho + \epsilon \Lambda_s^\theta \dd s},
	\end{equation}
	has a fixed point $W\in\II(h_{EZ},U)$. It is the unique fixed point such that $W\stackrel{\OO}{=}J$.
\end{prop}
\begin{rem}
	Often this theorem will be applied with $\Lambda=U$. In that case, if $U^\theta \in \bS\OO$ then there exists a fixed point $W$ of the operator $F_U=F^0_{U,U}$ and $W \stackrel{\OO}{=} J \stackrel{\OO}{=} U^\theta$.
\end{rem}
The following corollary to Proposition \ref{prop:comparison} shows that there is a natural ordering for subsolutions and supersolutions to \eqref{eq:perturbed fixed point operator} for different values of $\epsilon$ and $U$. In particular, since a solution is both a subsolution and a supersolution, it follows that for $\epsilon>0$, the solution found in Proposition \ref{prop:existence of a solution, bounded and perturbed} is the unique solution, and not just the unique solution of order $J$.
\begin{cor}\label{cor:comparison W,U positive eps} Fix $\nu>0$. Suppose $\Lambda\in\bS\OO_\nu^\theta$ and define the perturbed aggregator
	\begin{equation}\label{eq:perturbed aggregator}
		h^{\epsilon,\nu,\Lambda}_{EZ}(t,\omega,u,w)=u w^\rho + \epsilon e^{\nu t}\Lambda_t^\theta(\omega), \quad \text{for }\epsilon>0.
	\end{equation}
	Fix $\epsilon_2>0$ and $0\leq\epsilon_1\leq\epsilon_2$. Let $U^1,U^2\in\sP_+$ satisfy $U^1\leq U^2\leq \Lambda$. Suppose that $W^1$ is a subsolution for the pair $(h_{EZ}^{\epsilon_1,\nu,\Lambda},U^1)$ and $W^2$ is a supersolution for the pair $(h_{EZ}^{\epsilon_2,\nu,\Lambda},U^2)$ such that $W^1,W^2\in [0,\infty)$ and such that $(W^1,W^2)$ satisfies Condition~\ref{cond:V^1,V^2 condition} for the pair $(h_{EZ}^{\epsilon_1,\nu,\Lambda},U^1)$.
	Then, $W^1_\sigma \leq W^2_\sigma$ for all finite stopping times $\sigma\geq 0$.
\end{cor}
\begin{proof}
	First note that $W^2$ is a supersolution for $(h_{EZ}^{\epsilon_1,\nu,\Lambda},U^1)$, since
	\begin{equation}
		W^2_\sigma\geq\cEX[t]{\int_\sigma^\tau h_{EZ}^{\epsilon_2,\nu,\Lambda}(U^2_s,W^2_s)\dd s + W^2_{\tau+}}\geq\cEX[t]{\int_\sigma^\tau h_{EZ}^{\epsilon_1,\nu,\Lambda}(U^1_s,W^2_s)\dd s + W^2_{\tau+}}.
	\end{equation}
	As $\Lambda^\theta \in \bS\OO_\nu$, there exists $K_\Lambda$ such that $W^2_t \geq \cEX[t]{\int_t^\infty \epsilon_2 \e^{\nu s}\Lambda_s^\theta\dd s} \geq  \frac{\epsilon_2}{K_\Lambda}e^{\nu t}\Lambda_t^\theta$.	Therefore, since $U^1\leq \Lambda$ and $W^2 > 0$,
	\begin{equation}
		\dfrac{\partial h_{EZ}^{\epsilon_1,\nu,\Lambda}}{\partial w}\left(t,\omega,U^1_t,W^2_t\right) = \rho U^1_t(W^2_t)^{- \frac{1}{\theta}}  \leq\rho\left(\tfrac{K_\Lambda}{\epsilon_2}\right)^\frac{1}{\theta} e^{-\frac{\nu}{\theta} t}.
	\end{equation}
	Furthermore, $\frac{\partial^2 h_{EZ}^{\epsilon_1,\nu,\Lambda}}{\partial w^2}(u,w)=-\frac{\theta-1}{\theta^2}u w^{-(1+\frac{1}{\theta})}\leq 0$ for $w > 0$, so that $h_{EZ}^{\epsilon_1,\nu,\Lambda}$ is concave.
	Since $(W^1,W^2)$ satisfies Condition~\ref{cond:V^1,V^2 condition} for the pair $(h_{EZ}^{\epsilon_1,\nu,\Lambda},U^1)$, the assumptions of Proposition \ref{prop:comparison} are met, and the conclusion follows.
\end{proof}

The following Corollary gives explicit bounds $\hat{k},\hat{K}>0$ such that the fixed point $W$ found in Proposition \ref{prop:existence of a solution, bounded and perturbed} satisfies $\hat{k}J \leq W \leq\hat{K}J$.
\begin{cor}\label{cor:perturbed existence explicit bounds}
	Let $\epsilon\geq0$ and suppose that $U^\theta\in\bS\OO\big(k,K\big)$.  For $\epsilon>0$, suppose that $A$ and $B$ solve
	\begin{equation}\label{eq:A and B equation}
		A=K^{-1} (A^\rho + \epsilon) \qquad B=k^{-1} (B^\rho + \epsilon)
	\end{equation}
	and, if $\epsilon=0$, set $A = K^{-\theta}$ and $B = k^{-\theta}$ {\rm(}the positive solution to \eqref{eq:A and B equation}{\rm)}.
	Then, the fixed point $W$ of $F^\epsilon_{U,U}$ found in Proposition \ref{prop:existence of a solution, bounded and perturbed} is in $\OO(J; kA,KB)$.
\end{cor}
\begin{proof}
	We first show that $F^\epsilon_{U,U}$ maps from $\OO(U^\theta;A,B)$ to itself. We only prove the upper bound as the lower bound is symmetric.	Suppose that $W\leq BU^\theta$. Then,
	\begin{align*}
		F^\epsilon_{U,U}(W)_t =&~\cEX[t]{\int_t^\infty U_s W_s^\rho + \epsilon U_s^\theta \dd s}
		\\
		\leq&~ \cEX[t]{\int_t^\infty U_s  {B^\rho}U_s^{\theta\rho} + \epsilon U_s^\theta\dd s}
		\\
		=&~ \left( B^\rho + \epsilon\right)\cEX[t]{\int_t^\infty U_s^\theta \dd s}
		\\
		=&~ \left( B^\rho + \epsilon\right) J_t \leq \frac{1}{k} \left( B^\rho + \epsilon\right) U^\theta_t = B U^\theta_t.
	\end{align*}	
	The proof of Proposition \ref{prop:existence of a solution, bounded and perturbed} given in \cite[Theorem B.2]{herdegen2021infinite} first shows that $F^\epsilon_{U,U}: \OO(U^\theta)\to\OO(U^\theta)$ is a contraction mapping and then uses Banach's fixed point theorem. Hence, if we choose an initial process $W^0\in\OO(U^\theta;A,B)$, then repeated application of $F^\epsilon_{U,U}$ yields a fixed point $W^*\in\OO(U^\theta;A,B)$. Since the fixed point $W$ found in Proposition \ref{prop:existence of a solution, bounded and perturbed} is unique in the class $\OO(U^\theta)$, $W=W^*\in\OO(U^\theta;A,B)$. Finally, since $U^\theta\in\bS\OO(k,K)$, $\OO(U^\theta;A,B)\subseteq \OO(J;kA,KB)$.
\end{proof}

\section{The extremal solution}\label{sec:extremal solution}
In this section, we investigate the \textit{extremal} solution, from Definition \ref{def:extremal solution}. Recall that in the transformed coordinates the extremal solution is the maximal solution.
\begin{prop}\label{prop:maximal solution unique}
	If the maximal solution exists, it is unique.
\end{prop}
\begin{proof}
	
	Suppose for contradiction that there are two maximal solutions $W^1$ and $W^2$. Then, for all $t\geq0$, $W^1_t\geq W^2_t$ since $W^1$ is maximal in the class of solutions and $W^1_t\geq W^2_t$ since $W^1$ is maximal also. Thus, $W^1_t=W^2_t$ for all $t\geq0$. Since both $W^1$ and $W^2$ are c\`adl\`ag, they are indistinguishable.
\end{proof}
We now turn to existence of a maximal solution associated to $(h_{EZ},U)$.
\begin{prop}\label{prop:maximal existence U<K Lambda}
	Suppose that $U\in\sP_+$ satisfies $U\leq\Lambda$ for $\Lambda\in\bS\OO^\theta_+$. Then, there exists a unique maximal solution associated to $(h_{EZ},U)$. It is also maximal in the class of $L^1$-bounded subsolutions.
\end{prop}
\begin{proof}
	Since $\Lambda\in\bS\OO_+^\theta$, there exists $\nu>0$ such that $\Lambda\in\bS\OO_{\nu\theta}^\theta$. For such $\nu$, let $\Lambda^{(\nu)}_t = e^{\nu t}\Lambda_t$.
	Then, $\Lambda^{(\nu)}=(\Lambda^{(\nu)}_t)_{t\geq0}\in\bS\OO^\theta$. For each $n\in\N$, let $U^n \coloneqq \max\left\{U,\frac{1}{n}\Lambda^{(\nu)}\right\}$. Then, $U^n\stackrel{\OO}{=}\Lambda^{(\nu)}$ as $U^n\leq \Lambda\leq \Lambda^{(\nu)}$. Let $(\epsilon_n)_{n\in\N}$ be a positive-valued sequence such that $\epsilon_n\searrow0$.
	
	By Proposition \ref{prop:existence of a solution, bounded and perturbed}, for each $n\in\N$, there exists a solution $W^n$ associated to $(h_{EZ}^{\epsilon_n,\nu\theta,\Lambda},U^n)$, where $h_{EZ}^{\epsilon_n,\nu\theta,\Lambda}$ is defined in \eqref{eq:perturbed aggregator}. Furthermore, $W^n$ is decreasing in $n$ by Corollary \ref{cor:comparison W,U positive eps} and $U^n (W^n)^\rho$ is dominated by $U^1 (W^1)^\rho$. Hence, by the Dominated Convergence Theorem, we find that ${W \coloneqq \lim_{n\to\infty}W^n}$ satisfies
	\begin{equation}
		W_t = \lim_{n\to\infty}\cEX[t]{\int_t^\infty U^n_s (W^n_s)^\rho + \epsilon_n e^{\nu\theta s}\Lambda_s^\theta \dd s}=
		\cEX[t]{\int_t^\infty U_s W_s^\rho\dd s},
	\end{equation}
	so that $W\in\II(h_{EZ},U)$ is a solution associated to $(h_{EZ},U)$.
	
	Suppose that $W'\in\II(h_{EZ}, U)$ is a solution (or an $L^1$-bounded subsolution) associated to $(h_{EZ},U)$. Then, the pair $(W',W^n)$ satisfies Condition \ref{cond:V^1,V^2 condition} for the pair $(h_{EZ},U)$, since $W^n\in\UU\II(h_{EZ}^{\epsilon_n,\nu\theta,\Lambda},U^n)\subset\UU\II(h_{EZ},U)$ and $(W'-W^n)^+\leq W'$, where $W'$ is $L^1$-bounded. Hence, $W^n \geq W'$ for each $n\in\N$ by Corollary \ref{cor:comparison W,U positive eps} and $W \geq W'$ is a maximal ($L^1$-bounded sub-) solution. Uniqueness in the class of maximal solutions follows from Proposition \ref{prop:maximal solution unique}.
\end{proof}
\textit{Proposition \ref{prop:extremal solution for C^1-R < hatC^1-R}} is a direct result of the Proposition \ref{prop:maximal existence U<K Lambda} and the following comparison result for maximal solutions.

\begin{prop}\label{prop:maximal solutions increasing in U}
	Let $h_{EZ}$ be the aggregator defined in \eqref{eq:aggregator h_EZ} and suppose that $U^1,U^2\in\sP_+$ satisfy $U^1\leq U^2 \leq \Lambda\in\bS\OO^\theta_+$. If $W^1$ and $W^2$ are the maximal solutions associated to $h_{EZ}$ and consumption $U^1$ and $U^2$ respectively, then $W^1_t\leq W^2_t$ for all $t\geq0$.
\end{prop}
\begin{proof}
	Let $\nu$ be such that $\Lambda\in\bS\OO_{\nu\theta}^\theta$. Define $\Lambda^{(\nu)}=\big(\Lambda^{(\nu)}_t\big)_{t\geq0}$ by $\Lambda^{(\nu)}_t = e^{\nu t}\Lambda_t$ and for $n\in\N$ and $i\in\{1,2\}$ define $U^{i,n}=\max\big\{U^i,\frac{1}{n}\Lambda^{(\nu)}\big\}$ and ${\epsilon_n = \frac{1}{n}}$. Then, by Proposition \ref{prop:existence of a solution, bounded and perturbed}, there exists a solution $W^{i,n}$ associated to $(h_{EZ}^{\epsilon_n,\nu\theta,\Lambda},U^{i,n})$. Furthermore, for all $n\in\N$ and $t\geq0$, $W^{1,n} \leq W^{2,n}$ by Corollary \ref{cor:comparison W,U positive eps}.
	
	As in Proposition \ref{prop:maximal existence U<K Lambda}, the unique maximal solution associated to $U^i$ is given by $W^i \coloneqq \lim_{n \to \infty}W^{i,n}$ for $i\in\{1,2\}$. Thus, $W^1_t=\lim_{n \to \infty}W^{1,n}_t \leq \lim_{n \to \infty}W^{2,n}_t= W^2_t$ for all $t\geq0$.
\end{proof}
We may also deduce the following Corollary to Proposition  \ref{prop:maximal existence U<K Lambda}.
\begin{cor}\label{cor:extremal is maximal subsolution}
	Let $C\in\sP_+$ be such that $C^{1-R}\leq Y\in\bS\OO_+$. Then, the extremal solution associated to $(f_{EZ},C)$ is the maximal $L^1$-bounded subsolution when $R<1$ and the minimal $L^1$-bounded supersolution when $R<1$.
\end{cor}

The final result of this section shows that the solution found by fixed point argument in Proposition \ref{prop:existence of a solution, bounded and perturbed} is the unique maximal solution.
\begin{prop}\label{prop:fixed point solution is maximal solution for O(Lambda)}
	Let $h_{EZ}$ be the aggregator defined in \eqref{eq:aggregator h_EZ}. Suppose that $U\in\bS\OO^\theta_+$. Then the solution associated to $(h_{EZ},U)$ found in Proposition \ref{prop:existence of a solution, bounded and perturbed} is the maximal solution.
\end{prop}
\begin{proof}
	Fix $\theta>1$.	Since $U\in\bS\OO^\theta_+$, there exists $\hat{\nu}>0$ such that $U\in\bS\OO_{\hat{\nu}\theta}^\theta$. By Lemma \ref{lem:SO^mu subset SO^nu}, it further follows that $U\in\bS\OO_{\nu\theta}^\theta$ for $\nu\leq\hat{\nu}$. For each $\nu\leq\hat{\nu}$, define $U^{(\nu)}_t = e^{-\nu t}U_t$ and $J^{(\nu)}=(J^{(\nu)}_t)_{t\geq0}$ by $J^{(\nu)}_t = \cEX[t]{\int_t^\infty e^{\nu s} U^\theta_s \dd s}$. We can then find $k(\nu),K(\nu)$ such that $\left(U^{(\nu)}\right)^\theta\in\bS\OO(k(\nu),K(\nu))$. By Remark \ref{rem:bounds increasing in nu}, we may choose $k(\nu),K(\nu)$ such that $0<k(\hat{\nu})\leq\lim_{\nu\to0}k(\nu)=k(0)\eqqcolon k<\infty$ and $0<K(\hat{\nu})\leq\lim_{\nu\to0}K(\nu)=K(0)\eqqcolon K<\infty$, where both limits are decreasing in $\nu$.
	
	For each $\epsilon>0$ and $0<\nu\leq\hat{\nu}$, there exists a solution $W^{\epsilon,\nu}$ associated to $(h_{EZ}^{\epsilon,\nu\theta,\Lambda},U^{(\nu)})$ by Proposition \ref{prop:existence of a solution, bounded and perturbed}.
	Furthermore, by Corollary \ref{cor:perturbed existence explicit bounds}, $W^{\epsilon,\nu}_t\leq K(\nu) B^{\epsilon,\nu} J^{(\nu)}_t$ where $B=B^{\epsilon,\nu}$ solves $B=k(\nu)^{-1}(B^\rho + \epsilon)$. By Proposition \ref{prop:maximal existence U<K Lambda}, the unique maximal solution associated to $U$ is given by $W\coloneqq\lim_{\epsilon\to0}W^{\epsilon,\nu}$. Therefore, since $\lim_{\epsilon\to0}B^{\epsilon,\nu}=B^{0,\nu}=k(\nu)^{-\theta}$, $W_t \leq K(\nu)k(\nu)^{-\theta}J^{(\nu)}$ for all $\nu\leq \hat{\nu}$. Taking the limit as $\nu\searrow0$, gives $W_t\leq K k^{-\theta}J^{(0)}\eqqcolon K k^{-\theta}J$.
	
	Similarly, by maximality of $W$ and the lower bound found in Corollary \ref{cor:perturbed existence explicit bounds}, $W\geq kK^{-\theta} J$. Hence, the maximal solution is in $\OO(J)$. Since the solution found in Proposition \ref{prop:existence of a solution, bounded and perturbed} is unique in $\OO(J)$, it is equal to the maximal solution.
\end{proof}

\section{The proper solution}\label{sec:proper solution}

We first focus on existence of proper solutions and then turn to uniqueness.

\subsection{Existence of proper solutions}

The goal of this section is to prove Theorem \ref{thm:proper solution rc}. To this end, we first prove that there exists a proper solution $W$ associated to the aggregator $h_{EZ}$ and consumption stream $U$ given by a discounted indicator function of a stochastic interval, i.e. $U_t = e^{-\gamma t} \1_{\{\sigma \leq t < \tau\}}$
for $\sigma$ and $\tau$ stopping times such that $\sigma \leq \tau$. Since any right-continuous consumption stream can locally be approximated from below by (a scaled version of) these processes, we can show that there exists a proper solution associated to right-continuous processes.
\begin{prop}\label{prop:proper solution discounted 1_[sigma,tau)}
	Let $\gamma>0$ and $\sigma$ and $\tau$ be stopping times such that $\sigma\leq\tau$. Let ${U=(U_t)_{t\geq0}}$ be given by $U_t:=e^{-\gamma t}\1_{\{\sigma\leq t <\tau\}}$. Then, there exists a proper solution $W=(W_t)_{t\geq0}$ associated to $U$, for which $W_t \geq \big(\frac{1}{\gamma\theta}\cEX[t]{e^{-\gamma (t\vee\sigma)}- e^{-\gamma(t\vee\tau)}}\big)^\theta$.
\end{prop}
The proof of Proposition \ref{prop:proper solution discounted 1_[sigma,tau)} is long and technical and therefore relegated to the appendix.

To prove Theorem \ref{thm:proper solution rc}, we introduce two technical lemmas. 
\begin{lemma}\label{lemma:utility process associated to 1_A U}
	Suppose that $W$ is a utility process associated to $(h_{EZ},U)$ and suppose that there exists $t_0\geq0$ such that $U_t=0$ for $t<t_0$. If $A\in\cF_{t_0}$, and $\widetilde{U}=(\widetilde{U}_t)_{t\geq0}$ is given by $\widetilde{U}_t = \cEX[t]{\1_A}U_t$ then $\widetilde{W}_t = \cEX[t]{\1_AW_{t\vee t_0}}$ is a utility process associated to $(h_{EZ},(\widetilde{U}_t)_{t\geq0})$.
\end{lemma}
\begin{proof}
	Suppose that $t\geq t_0$. Then, since $W$ is a utility process associated to $U$,
	\begin{align}
		\widetilde{W}_t = \1_AW_t = \1_A^{1+\rho}W_t
		= \cEX[t]{\int_t^\infty  \1_AU_s (\1_AW_s)^\rho\dd s}
		= \cEX[t]{\int_t^\infty  \widetilde{U}_s \widetilde{W}_s^\rho\dd s}.
	\end{align}
	Conversely, suppose that $t<t_0$. Then, since $\widetilde{U}_s=0$ for $s<t_0$ and both $\widetilde{W}_{t_0} = \1_A W_{t_0}$ and $\widetilde{W}_{t_0} = \cEX[t_0]{\int_{t_0}^\infty \widetilde{U}_s \widetilde{W}_s^\rho\dd s}$,
	\begin{equation*}
		\widetilde{W}_t = \cEX[t]{\1_A W_{t_0}}=
		\cEX[t]{ \int_{t_0}^\infty \widetilde{U}_s \widetilde{W}_s^\rho\dd s} =\cEX[t]{\int_t^\infty \widetilde{U}_s \widetilde{W}_s^\rho\dd s}.\qedhere
	\end{equation*}
\end{proof}
\begin{lemma}\label{lemma:J^U,theta>0 implies U_T>eps for T>t}
	Suppose $U\in\sP_+$ is right-continuous. Fix $t \geq 0$ and let
	${A_t=\{J^{U^\theta}_t > 0\}}$ and $B_t=\bigcup_{\substack{T\geq t\\T\in\QQ}}\bigcup_{\substack{\epsilon>0\\ \epsilon\in\QQ}} \left\{\cEX[t]{\1_{\{U_T\geq\epsilon\}}}>0\right\}$. Then, $\P(A_t\setminus B_t)=0$.
\end{lemma}
\begin{proof}
	Seeking a contradiction, suppose $C_t\coloneqq A_t\setminus B_t$  has positive measure. Then $\E[\1_{C_t} J^U_t]>0$  by the definition of $A_t$. Moreover, for each \textit{rational} $T \geq t$, the definition of $B_t$ gives $\cEX[t]{\1_{C_t} U_T}\leq\cEX[t]{\1_{B_t^c} U_T}=0$ which yields $\1_{C_t} U_T = 0$ $\as{\P}$~ Since $U$ is right-continuous, $\1_{C_t} U_T=0$ for all $T\geq t$ $\as{\P}$~ Taking expectations yields $\E[\1_{C_t} J^U_t] = \EX{\int_t^\infty \1_{C_s} U_s \dd s}=0$ and we arrive at a contradiction.
\end{proof}
We may now prove Theorem \ref{thm:proper solution rc}. To show that $W$ is proper, we must show that if $A_t=\{J^{U^\theta}_t > 0\}$ and $C_t =\{W_t>0\}$, then $\P(A_t\setminus C_t)=0$. By Lemma \ref{lemma:J^U,theta>0 implies U_T>eps for T>t}, since $A_t\subseteq B_t$ up to null sets, we may instead prove that $\P(B_t\setminus C_t)=0$.
\begin{proof}[Proof of Theorem \ref{thm:proper solution rc}]\label{proof:thm:proper solution rc process}
	Since $C^{1-R}\leq Y$ for $Y\in\bS\OO_+$, $U\coloneqq\theta C^{1-S}\leq\theta Y^{\frac{1}{\theta}}$. As ${Y\in\bS\OO_+}$ and by Remark \ref{rem:YinSO implies KYinSO}, $\theta Y^{\frac{1}{\theta}}\in\bS\OO^\theta_+$.
	Therefore, by Proposition \ref{prop:maximal existence U<K Lambda} there exists a maximal solution $W$ associated to $(h_{EZ},U)$. We now show that $W$ is proper. 
	
	Fix $t^* \geq 0$. Set ${A_{t^*}:=\{J^{U^\theta}_{t^*}  > 0\}}$ and $C_{t^*} =\{W_{t^*}>0\}$. By Lemma~\ref{lemma:J^U,theta>0 implies U_T>eps for T>t}, it suffices to show that $\P(B^\epsilon_T \setminus C_{t^*}) = 0$ for all rational $T \geq {t^*}, \epsilon > 0$, where $B^\epsilon_T=\{\cEX[{t^*}]{\1_{\{U_T\geq\epsilon\}}}>0\}$. So, fix rational $T\geq {t^*}$ and $\epsilon > 0$. Define the stopping time $\tau_\epsilon=\inf\{t\geq T: U_t\leq \frac{\epsilon}{2}\}$ and note that $\{\tau_\epsilon > T\}$ on $\{U_T \geq\epsilon\}$ by right-continuity of $U$. Define the process $\widetilde{U} = (\widetilde{U}_t)_{t \geq 0}$ by
	${\widetilde{U}_t := \frac{\epsilon}{2} e^{-\gamma t}\cEX[t]{\1_{\{U_T\geq\epsilon\}}} \1_{\{t\in[T,\tau_\epsilon)\}}}$. Then $\widetilde{U}$ is dominated by $U$. Moreover, Proposition \ref{prop:proper solution discounted 1_[sigma,tau)} for $\hat U_t = e^{-\gamma t} \1_{\{T \leq t < \tau_\epsilon\}}$ with corresponding solution $\hat W$, Lemma \ref{lemma:utility process associated to 1_A U} for $t_0=T$ and $A = \{ \hat U_T \geq \epsilon \}$ and Jensen's inequality show that there exists a solution $\widetilde{W}$ associated to $\widetilde{U}=(\widetilde{U}_t)_{t\geq0}$ such that for all $t \geq 0$
	\begin{align}
		\widetilde{W}_t \geq&~\cEX[t]{\1_{\{U_T\geq\epsilon\}}\left(\frac{\epsilon}{2\gamma\theta}\cEX[t]{(e^{-\gamma (t\vee T)}- e^{-\gamma(t\vee\tau_\epsilon)})}\right)^\theta}
		\\
		\geq&~\left(\frac{\epsilon}{2\gamma\theta}\cEX[t]{\1_{\{U_T\geq\epsilon\}}}\cEX[t]{(e^{-\gamma (t\vee T)}- e^{-\gamma(t\vee\tau_\epsilon)})}\right)^\theta.
	\end{align}
	Since $\{\tau_\epsilon > T\}$ on $\{U_T \geq\epsilon\}$, it follows that $\tilde W_{t^*} > 0$ on $\{U_T \geq\epsilon\}$. Now the claim follows from the fact that $W_{t^*}\geq\widetilde{W}_{t^*}$ by Proposition \ref{prop:maximal solutions increasing in U}.
\end{proof}

\subsection{Uniqueness of proper solutions}
We now turn to uniqueness of proper solutions. The aim of this section will be to prove Theorem \ref{thm:O^1-R_+ subset UP} and then, as a corollary, Theorem \ref{thm:three solutions coincide}. The following two lemmas will be useful.
\begin{lemma}\label{lem:J = Me^A}
	Fix $X \in \bS\OO$, and let $J^X=(J^X_t)_{t\geq0}$ be defined by $J^X_t = \cEX[t]{\int_t^\infty X_s ds}$. Then, there exists a martingale $M=(M_t)_{t\geq0}$ such that $J^X_t = M_t e^{- \int_0^t (X_s/J^X_s) ds }$.
\end{lemma}
\begin{proof}
	Let $N=(N_t)_{t\geq0}$ be the uniformly integrable martingale given by $N_t = \cEX[t]{\int_0^\infty X_s \dd s}$. Then $J^X_t = N_t - \int_0^t X_s \dd s$. Define the increasing process $A$ by $A_t = \int_0^t (X_s/J^X_s) \dd s$. Since $X \in \bS\OO$ we have $0< X_t \leq K J^X_t$ for some $K$ and hence $0 < A_t \leq Kt$.
	
	Define $M$ via $M_t = e^{A_t} J^X_t$. Then $dM_t = e^{A_t} dJ^X_t + X_t e^{A_t} dt = e^{A_t} dN_t$ and all that remains to show is that the local martingale $M$ is a martingale. Since $A$ is increasing and $A_t \leq Kt$, we have $\E[\lVert e^{A}\rVert_T|N_T|]\leq\E[(e^{KT}-1)|N_T|]<\infty$ for $T\geq0$, where $\lVert e^{A}\rVert_T$ is the total variation of $e^A=(e^{A_t})_{t\geq0}$ at time $T$. Hence, $M$ is a martingale by \cite[Lemma A.1]{herdegen2019sensitivity}.
\end{proof}

\begin{lemma}\label{lemma:X>1 finite time}
	Let $\alpha>0$, $\beta\in(0,1)$ and let $G=(G_t)_{t\geq0}$ be a c\`adl\`ag submartingale. Suppose that $X=(X_t)_{t\geq0}$ is is a right-continuous process such that $X_0=\beta$ and $X_t\leq1$ for all $t\geq0$. Define $\tau=\inf\{t\geq0:~X_t=1\}$ and suppose that
	\begin{equation}\label{eq:Xdynam}
		\dd X_t \geq \alpha X_t \dd t + \dd G_t, \quad\text{for all }t<\tau.
	\end{equation}
	Then, there exists $\nu \in (0,1)$ and $T \in (0,\infty)$ such that $\P(\tau<T)>\nu$.
\end{lemma}
\begin{proof}
	First note that $X$ is a (local) submartingale bounded above by 1 and so converges almost surely to an $\cF_\infty$-measurable random variable $X_\infty\leq1$ by the Martingale Convergence Theorem.	
	
	Fix $\xi\in(0,\beta)$. Let $\sigma=\inf\{t\geq0:~X_t\notin(\xi,1)\}\leq\tau$. From the dynamics of $X$ given in \eqref{eq:Xdynam},
	$$X_{t\wedge\sigma} \geq X_0 +\int_0^{t\wedge\sigma} \alpha X_s \dd s + G_{t\wedge\sigma}-G_0 \geq \beta+\alpha\xi (t\wedge\sigma) + G_{t\wedge\sigma}-G_0.$$
	Then, using that $G$ is a c\`adl\`ag submartingale and the Optional Sampling Theorem,
	\begin{equation}\label{eq:stoppped X inequality}
		\EX{X_{t\wedge\sigma}}\geq \beta+\alpha\xi \EX{t\wedge\sigma}.
	\end{equation}
	Since $X\leq1$, taking the $\limsup$ and using the Reverse Fatou's Lemma on the left hand side of \eqref{eq:stoppped X inequality} and the Monotone Convergence Theorem on the right hand side gives
	\begin{equation}
		1\geq\EX{X_\sigma}=\EX{\1_{\sigma<\infty}X_\sigma}+\EX{\1_{\sigma=\infty}X_\infty}\geq \beta+\alpha\xi \EX{\sigma}\geq \beta.
	\end{equation}
	Therefore, $\E[\sigma] \leq \frac{1-\beta}{\alpha \xi}$ and $\P(\sigma=\infty)=0$
	. Consequently, since $X$ is right-continuous, $X_\sigma\in(-\infty,\xi]\cup\{1\}$ $\as{\P}$ and
	\begin{equation}
		1 - (1-\xi)\P(X_\sigma\leq\xi)=\P(X_\sigma=1)+\xi\P(X_\sigma\leq\xi)\geq\EX{X_\sigma}\geq \beta.
	\end{equation}
	In particular, $\P(X_\sigma\leq\xi)\leq \frac{1-\beta}{1-\xi}$ and $\P(\sigma=\tau)=\P(X_\sigma=1)\geq 1- \frac{1-\beta}{1-\xi} = \frac{\beta-\xi}{1-\xi}$. Furthermore, 
	\begin{align}
		\P(\sigma\geq T; \sigma=\tau)\leq \EX{\frac{\sigma}{T}\1_{\sigma\geq T}\1_{\sigma=\tau}}\leq \frac{1}{T}\EX{\sigma}\leq \frac{1-\beta}{\alpha\xi T}, \quad \text{for all }T\geq0,
		\shortintertext{and}
		\P(\tau<T)\geq\P(\sigma<T;\sigma=\tau)=\P(\sigma=\tau)-\P(\sigma\geq T;\sigma=\tau)\geq\frac{\beta-\xi}{1-\xi} - \frac{1-\beta}{\alpha\xi T}.
	\end{align}
	
	Choose $\nu = \frac{1}{2}\left(\frac{\beta-\xi}{1-\xi}\right)$ and $T = \frac{1-\beta}{\alpha\xi \nu}$. Then, $\P(\tau<T)\geq \nu$.
\end{proof}

We may now prove Theorem \ref{thm:O^1-R_+ subset UP} in the $(U,W)$ coordinates.

\begin{proof}[Proof of Theorem \ref{thm:O^1-R_+ subset UP}]
	Fix $U \in\bS\OO^\theta_+$ and recall $J=(J_t)_{t\geq0}$ is given by $J_t = \cEX[t]{\int_t^\infty U^\theta_s \dd s}$. Suppose that $U^\theta\in \bS\OO(k,K)$ and let $W\in\OO(J)$ be the solution associated to $(h_{EZ},U)$ found in Proposition \ref{prop:existence of a solution, bounded and perturbed} (after setting $\epsilon = 0$ and $U = \Lambda$). By Proposition \ref{prop:fixed point solution is maximal solution for O(Lambda)}, $W$ is the unique maximal solution, and as we saw in the proof of Theorem~\ref{thm:proper solution rc} it is also proper. We now prove uniqueness.
	
	For contradiction, assume that there exists a proper solution $Y=(Y_t)_{t\geq0}$ such that $Y\neq W$. Since $W$ is maximal, $Y\leq W$. Then, since $W$ is unique in the class $\OO(J)$, it follows that $Y\stackrel{\OO}{\neq}J$. Hence, there exists $t\geq0$, $B\in\cF_t$ and $\epsilon>0$ such that $\P(B)>\epsilon$ and $Y_t<kK^{-\theta} J_t$ on $B$.	For ease of exposition, assume that $t=0$, $B=\Omega$ and $Y_0=y=(1-\epsilon)kK^{-\theta}J_0$. The general case is similar.
	
	Define $Z=(Z_t)_{t\geq0}$ by $Z_t=Y_t(kK^{-\theta}J_t)^{-1}$ and note that $Z$ is c\`adl\`ag by Remark \ref{rem:solutions are utility processes} and $Z_0=(1-\epsilon)$. Since $kK^{-\theta}JZ \equiv Y$ is a utility process and since $U^\theta\geq kJ$, for all $t\leq T$,
	\begin{align}
		kK^{-\theta}J_t Z_t = Y_t =~ &\cEX[t]{\int_t^T U_s Y_s^\rho \dd s+ Y_T}
		\\
		\geq~&\cEX[t]{\int_t^T (kJ_s)^\frac{1}{\theta} (kK^{-\theta}J_sZ_s)^\rho \dd s+kK^{-\theta}J_T Z_T}.
		\intertext{By Lemma \ref{lem:J = Me^A}, we find that $J_t = M_t e^{-A_t}$ for $A_t = \int_0^t (U_s^\theta/J_s) ds$. Hence, dividing by $kK^{-\theta}M_t$ and collecting terms gives}
		e^{-A_t}Z_t \geq~ \frac{1}{M_t}&\cEX[t]{\int_t^T K M_s e^{-A_s}Z_s^\rho \dd s+M_T e^{-A_T} Z_T}.
	\end{align}	
	Define $\widetilde{Z}_t=e^{-A_t}Z_t$ and consider an equivalent measure $\widetilde{\P}$ defined by $\frac{\mathrm{d}\widetilde{\P}}{\mathrm{d} \P}\big|_{\cF_t} = M_t$, so that
	\begin{equation}\label{eq:Zeq first}
		\widetilde{Z}_t \geq \cEXt[t]{\int_t^T K e^{-(1-\rho)A_s} \widetilde{Z}_s^\rho \dd s +\widetilde{Z}_T}, \quad \text{for all }t\leq T.
	\end{equation}
	If we define $O$ by $O_t = \widetilde{Z}_t + \int_0^t K e^{-(1-\rho)A_s} \widetilde{Z}_s^\rho \dd s$, then $O$ is a c\`adl\`ag supermartingale. By the Doob--Meyer decomposition, $O$ can therefore be decomposed as $O = N + P$, where $N$ is a local martingale and $P$ is a decreasing process such that $P_0=0$, both of which are c\`adl\`ag. In particular, rearranging gives
	\begin{equation}\label{eq:subsolution as subsolution to BSDE?}
		\dd\widetilde{Z}_t =  - K e^{-(1-\rho)A_t}\widetilde{Z}_t^\rho \dd t + \dd N_t -\dd P_t.
	\end{equation}
	
	Let $\sigma\coloneqq \{t\geq0: Z_t\notin(0,1)\}$, $\hat{Z}_t=Z_{t\wedge{\sigma}}$, $\hat{N}_t = \int_0^{t\wedge\sigma}e^{A_s}\dd N_s$ and $\hat{P}_t=\int_0^{t\wedge \sigma} e^{A_s}\dd P_s$. Then, applying the product rule to $\hat{Z}_t=e^{A_t}\widetilde{Z}_t$ up to $t\leq \sigma$, and noting that $\frac{\dd A_t}{\dd t} = \frac{U_t^\theta}{J_t}\leq K$,
	\[		\dd \hat{Z}_t = \hat{Z}_t\dd A_t - K\hat{Z}_t^\rho \dd t+ \dd \hat{N}_t - \dd \hat{P}_t
	\leq K(\hat{Z}_t - \hat{Z}_t^\rho)\dd t + \dd \hat{N}_t \leq \dd \hat{N}_t.\]
	Since $\hat{N}_t \geq \hat{Z}_t - \hat{Z}_0 \geq - \hat{Z}_0$, $\hat{N}$ is a supermartingale.

	Let $X_t=1-\hat{Z}_t^{1-\rho}\leq 1$. Then, $X=(X_t)_{t\geq0}$ is c\`adl\`ag and, for $t<\sigma$,
	\begin{align}
		\dd X_t =&~ -\!{(1-\rho)}\hat{Z}_t^{-\rho}\dd \hat{Z}_t + \frac{\rho(1-\rho)}{2}\hat{Z}_t^{-(\rho+1)}\dd \langle Z\rangle_t
		\\
		=&~ -\!{(1-\rho)}(\hat{Z}_t^{1-\rho}\dd A_t -K\dd t) + \dd L_t + \dd Q_t
		\\
		\geq&~ K(1-\rho)X_t \dd t + \dd L_t + \dd Q_t \geq \dd L_t
	\end{align}
	where
	\begin{equation}
		L_t \coloneqq -{(1-\rho)}\!\!\int_0^t \!\hat{Z}^{-\rho}_s\dd \hat{N}_s \quad \text{and} \quad Q_t\coloneqq(1-\rho)\!\!\int_0^t\!\hat{Z}^{-\rho}_s\dd \hat{P}_s + \int_0^t \frac{\rho(1-\rho)}{2}\hat{Z}^{-(\rho+1)}\dd \langle Z\rangle_s.
	\end{equation}
	Since $L_t \leq X_t - X_0 \leq 1$, $L$ is a (continuous) submartingale. Hence, $G \coloneqq L+Q$ is a continuous submartingale. The result that $X$ explodes to $1$ in finite time with positive probability follows from Lemma \ref{lemma:X>1 finite time}. This implies that $Z$ hits zero in finite time and, consequently, that $Y$ is not proper.
\end{proof}
\begin{proof}[Proof of Theorem \ref{thm:three solutions coincide}]
	Let $W$ be the unique solution such that $W\stackrel{\OO}{=}J$ given by Proposition \ref{prop:O solution for C^1-R in SO} ($V=\frac{W}{1-R}$ is the CRRA-order solution). Since $J>0$, $W$ is proper and uniqueness in the class of proper solutions follows from Theorem \ref{thm:O^1-R_+ subset UP}. Finally, $W$ is the maximal solution by Proposition \ref{prop:fixed point solution is maximal solution for O(Lambda)}.
\end{proof}

\section{Verification of the candidate optimal strategy}\label{sec:verification}
This section aims to prove that the candidate optimal strategy given in Proposition \ref{prop:derivecandidate} is indeed optimal. We will roughly follow the approach detailed in \cite{herdegen2021infinite} which goes as follows: first, show that if $\hat{X}=X^{\hat{\Pi},\hat{C}}$ is the wealth process under the candidate optimal strategy, then $\hat{V}(X+\epsilon\hat{X})$ is a supersolution for $(f_{EZ},C)$  (The reasons that we perturb the input of $\hat{V}=\hat{V}(X)$ by the optimal wealth process are to ensure that we can apply It\^o's lemma to $\hat{V}$, to make sure that the local martingale part of $\hat{V}$ is a supermartingale, and to ensure that $\hat{V}$ satisfies the supersolution transversality condition. This is explained in more detail in \cite{herdegen2020elementary} and \cite[Section~11]{herdegen2021infinite}); next, use a version of the Comparison Theorem (Corollary \ref{cor:comparison C^1-R < K (1-R)V}) for sub- and supersolutions to conclude that $\hat V(x(1+\epsilon)) \geq V^C_0$; finally, letting $\epsilon\searrow0$ gives $\hat V(x) \geq V^C_0$. Optimality follows since $V^{\hat{C}}_0 = \hat V (x)$ by Proposition \ref{prop:derivecandidate}.

However, the approach is not quite this simple for two main reasons. The most pressing reason is that when $\theta>1$ the utility process $V^C$ fails to be unique. The optimisation therefore takes place over the attainable and right-continuous consumption streams for which there exists a unique proper solution to the EZ-SDU equation. As we have argued in Section~\ref{sec:proper solution}, the proper solution is the economically meaningful solution, and we may only consider the consumption streams that have a unique proper solution associated to them. The assumption of right-continuity, which is necessary for the proof, is not overly restrictive.

The next issue is that the hypotheses of the relevant comparison theorem (Corollary \ref{cor:comparison C^1-R < K (1-R)V}) are not satisfied, even for right-continuous consumption streams with a unique proper solution. To overcome this issue, one must approximate an arbitrary consumption stream in $\UU\PP^*$ by a series of consumption streams satisfying the conditions and then take limits. The requirement of right-continuity ensures that we may choose right-continuous approximating consumption streams, which then have an associated proper solution $V^n$ by Theorem \ref{thm:proper solution rc}. Since the limiting process $V=\lim_{n\to\infty}V^n$ is a proper solution associated to $C$, it must agree with the unique proper solution $V^C$.

To prove Theorem \ref{thm:verification}, we will use the following lemma, which is proved as an intermediate part of \cite[Theorem 11.1]{herdegen2021infinite}.  The result in \cite{herdegen2021infinite} is written for the case $0<\theta<1$, but it is not difficult to check that the argument extends to the case $\theta > 1$.
\begin{lemma}\label{lem:V hat (X+eps Y) a supersolution}
	Let $\epsilon>0$ and let $\hat{X}=X^{\hat{\Pi},\hat{C}}$ denote the wealth process under our candidate optimal strategy. Fix $(\Pi,C)$ and let $X = X^{\Pi,C}$ denote the corresponding wealth process. Then, $\hat{V}(X + \epsilon \hat{X})$ is a supersolution for the pair $(f_{EZ},C+\eta\epsilon \hat{X})$.
\end{lemma}
We may then prove Theorem \ref{thm:verification}. Note that $\hat{V}(\hat{X})$ is a solution for the pair $(f_{EZ}, \eta \hat{X})$ and, by scaling, $\hat{V}( \epsilon \hat{X})$ is a solution for the pair $(f_{EZ},\eta\epsilon\hat{X})$. We expect that $\hat{V}(X^{\Pi,C})$ is a supersolution for $(f_{EZ},C)$ but, when $R>1$, the transversality condition might not hold. Furthermore, the conditions required for Proposition \ref{prop:comparison} to hold may be impossible to verify. However, as we show in the proof below, by considering the perturbed problem, the transversality condition is guaranteed and the comparison theorem can be applied. 
\begin{proof}[Proof of Theorem \ref{thm:verification}]
	It follows from Proposition \ref{prop:derivecandidate} that $\hat{V}(\hat{X})$ is a utility process associated to candidate optimal strategy $(\hat{\Pi}, \hat{C})$. Since $\hat{V}(\hat{X})$ is a CRRA-order solution to \eqref{eq:Epstein--Zin SDU}, it is the unique proper solution to \eqref{eq:Epstein--Zin SDU} by Theorem \ref{thm:three solutions coincide}. Hence, $V^{\hat{C}}_0 = \hat{V}(x)$. It therefore only remains to show that $V^C_0\leq\hat{V}(x)$ for all $C \in \sC(x)\cap\UU\PP^*$. Fix an arbitrary $C \in \sC(x)\cap\UU\PP^*$ and let $\Pi=(\Pi_t)_{t\geq0}$ be an associated investment process.
	
	We first prove the result when $R>S>1$. In this case $V^C$ is the minimal solution (since ${C\in\UU\PP}$ and the minimal solution is proper).  By Lemma \ref{lem:V hat (X+eps Y) a supersolution}, for each $\epsilon>0$, $\hat{V}(X^{C,\Pi} + \epsilon \hat{X})$ is a supersolution associated to $C^\epsilon=C+\eta\epsilon\hat{X}$.
	Since $(C^\epsilon)^{1-S}\leq( \eta\epsilon)^{1-S}\hat{X}^{1-S}$, there exists an extremal solution $V^{C^\epsilon}$ associated to $C^\epsilon$ by Proposition \ref{prop:extremal solution for C^1-R < hatC^1-R} which is increasing in $\epsilon$ by Proposition \ref{prop:maximal solutions increasing in U}. It is the minimal supersolution by Corollary \ref{cor:extremal is maximal subsolution}. Hence, by minimality, $V^{C^\epsilon}_t\leq\hat{V}(X^{C,\Pi}_t + \epsilon \hat{X}_t)<0$ for all $t\geq0$ and $V^{C^\epsilon}$ is proper.

	Let $V^*=\lim_{\epsilon\to0}V^{C^\epsilon}$. Then, $V^*_0\leq\hat{V}(x)$. Consequently, since $f_{EZ}$ is increasing in both arguments, and $C^\epsilon$ and $V^\epsilon$ are increasing in $\epsilon$, $f_{EZ}(C^\epsilon,V^\epsilon)$ is increasing in~$\epsilon$, and applying the Monotone Convergence Theorem for conditional expectations yields
	\begin{equation*}
		V^*_t = \lim_{\epsilon\to0}V^{C^\epsilon}_t = \lim_{\epsilon\to0}\cEX[t]{\int_t^\infty f_{EZ}(C^\epsilon_s,V^{C^\epsilon}_s) \dd s} = \cEX[t]{\int_t^\infty f_{EZ}(C_s,V^*_s) \dd s}.
	\end{equation*}
	Therefore, $V^*$ is a solution associated to $(f_{EZ},C)$. Since $V^*_t = \lim_{\epsilon\to0}V^{C^\epsilon}_t<0$ for all $t\geq0$, $V^*$ is proper. It therefore agrees with the unique proper solution $V^C$ so that $V^C_0\leq\hat{V}(x)$.\\

	We now prove the result when $R<S<1$. Fix an arbitrary $C\in\sC(x)\cap\UU\PP^*$ with associated investment process $\Pi=(\Pi_t)_{t\geq0}$.	
	Let $0<\zeta<\eta\frac{S}{1-S}$ and define $\widetilde{X}_t=e^{\zeta t}X^{C,\Pi}_t$, $Y_t = e^{\frac{\zeta}{S}t}\hat{X}_t$ and $\widetilde{C}_t=e^{\zeta t}C_t$ for $t\geq0$. Note that if ${r_\zeta=r+\zeta}$ and $\mu_\zeta=\mu+\zeta$, then
	\begin{eqnarray*}
		\dd \widetilde{X}_t	&=& \widetilde{X}_t \Pi_t  \sigma \dd B_t + \left( \widetilde{X}_t (r_\zeta + \Pi_t (\mu_\zeta- r_\zeta)) - \widetilde{C}_t\right)\mathrm{d} t
	\end{eqnarray*}
	We may think of $\widetilde{X}=(\widetilde{X}_t)_{t\geq0}$ as being the wealth process associated to the strategy $(\Pi,\widetilde{C}=(\widetilde{C}_t)_{t\geq0})$ in a more favourable financial market with risk-free rate $r_\zeta$, drift of the risky asset $\mu_\zeta$, and well-posedness parameter $\eta_\zeta = -\frac{1-S}{S}(r_\zeta + \frac{\lambda^2}{2R}) = \eta - \frac{1-S}{S}\zeta \in (0,\eta)$. The volatility is unchanged. 
	Furthermore, since
	\begin{equation}
		\frac{dY_t}{Y_t } = \frac{\lambda}{R} \dd B_t + \left(\left(r + \frac{\lambda^2}{R} - \eta\right) + \frac{\zeta}{S}\right)\dd t = \frac{\lambda}{R} \dd B_t + \left( r_\zeta + \frac{\lambda^2}{R} - \eta_\zeta \right)\dd t,
	\end{equation}
	$Y=(Y_t)_{t\geq0}$ is the wealth process under the optimal strategy in the new financial market with parameters $r_\zeta$ and $\mu_\zeta$. Define $\hat{V}^\zeta(x) = \eta_\zeta^{-\theta S }\frac{x^{1-R}}{1-R}$. Then, $\hat{V}^\zeta(\widetilde{X}+\epsilon Y)$ is a supersolution for $(f_{EZ},\widetilde{C}+\eta \epsilon Y)$ by Lemma \ref{lem:V hat (X+eps Y) a supersolution} and then also for $(f_{EZ},C)$ since $\widetilde{C}+\eta\epsilon Y\geq C$.
	
	Let $C^n\coloneqq C\wedge n\hat{X}$. Then, there exists an extremal solution $V^{C^n}$ associated to $C^n$ by Proposition \ref{prop:extremal solution for C^1-R < hatC^1-R} which is monotonously increasing in $n$ by Proposition \ref{prop:maximal solutions increasing in U}. Also, $\hat{V}^\zeta = \hat{V}^\zeta(\widetilde{X}+\epsilon Y)$ is a supersolution for $(f_{EZ},C^n)$ since $C\geq C^n$. Furthermore, since $C$ is right-continuous, $C^n$ is right-continuous. The extremal solution $V^{C^n}$ associated to $C^n$ is therefore proper by Theorem \ref{thm:proper solution rc} and Remark \ref{rem:maximal solution proper}.
	
	Next, using that $(C^n)^{1-R}\leq n^{1-R}\hat{X}^{1-R}$, we obtain
	\begin{align}
		(1-R)\hat{V}^\zeta(\widetilde{X}+\epsilon Y) &\geq \eta_\zeta^{-\theta S }(\epsilon Y)^{1-R} = \eta_\zeta^{-\theta S }\epsilon^{1-R} e^{\frac{\zeta(1-R)}{S}t}\hat{X}^{1-R} \\
		&\geq \frac{\eta_\zeta^{-\theta S }\epsilon^{1-R} }{n^{1-R}}e^{\frac{\zeta(1-R)}{S}t} (C^n)^{1-R}.
	\end{align}
	Furthermore, 
	$V^{C^n}\in\UU\II(f_{EZ},C^n)$ by Remark \ref{rem:utility process UI} and $$\EX{(V^{C^n}_t-\hat{V}^\zeta_t)^+}\leq \EX{V^{C^n}_t}=\EX{\int_t^\infty f_{EZ}(C^n_s,V^{C^n}_s)\dd s}\leq \EX{\int_0^\infty f_{EZ}(C^n_s,V^{C^n}_s)\dd s}<\infty.$$ Hence, the conditions of Corollary \ref{cor:comparison C^1-R < K (1-R)V} are met and $\hat{V}^\zeta_t \geq V^{C^n}_t$ for all $t\geq0$. In particular, $\hat{V}^\zeta_0 \geq V^{C^n}_0$. If $V^*=\lim_{n\to\infty}V^{C^n}$ is the monotone limit, then
	\begin{equation}
		V^*_t = \lim_{n\to\infty}V^{C^n}_t = \lim_{n\to\infty}\cEX[t]{\int_t^\infty f_{EZ}(C^n_s, V^{C^n}_s)\dd s}=\cEX[t]{\int_t^\infty f_{EZ}(C_s, V^*_s)\dd s}.
	\end{equation}
	Hence, $V^*$ is a solution. It is a proper solution since for each $t \geq 0$, $V^*_t > 0$ if $V^{C^n}_t>0$ for some $n$ and $V^{C^n}_t>0$ on $\{J^{(C^n)^{1-R}} > 0\} = \{J^{C^{1-R}} > 0\}$ up to null sets by the fact that each $V^{C^n}$ is proper and $\hat X$ is strictly positive. Therefore, $V^*$ must agree with the unique proper solution $V^C$ associated to $C$. In particular, $\hat{V}^\zeta_0=\hat{V}^\zeta(x(1+\epsilon))\geq \lim_{n\to\infty} V^{C^n}_0 =V^*_0 = V^C_0$. Finally, taking $\zeta,\epsilon\searrow0$ gives $\hat{V}(x)\geq V^C_0$.
\end{proof}
	
	\bibliographystyle{plain}
	\bibliography{proper_solutions}
	{\small
	\include{bibliography}
}
	\appendix
	
	\section{Existence of proper solutions associated to discounted indicators of stochastic intervals}
	\label{sec:proofs existence of proper}
	
	This section will be dedicated to proving Proposition \ref{prop:proper solution discounted 1_[sigma,tau)}. In this section, we will emphasise the role that the filtration $\FF=(\cF_t)_{t\geq0}$ plays in determining the utility process $V=(V_t)_{t\geq0}$ associated to a pair $(g,C)$. To this end, if $\FF=(\cF_t)_{t\geq0}$ is the filtration used in Definition \ref{defn:integrable set I(g,C)}, then we will refer to $V$ as the utility process associated to \textit{the triple} $(g,C,\FF)$.
	
	Let $\FF=(\cF_t)_{t\geq0}$ be a filtration and $\TT=\{t_0,t_1,t_2,\cdots,t_n\}$ be an ordered set. We assume without loss of generality that $t_0=0$ and $t_n=\infty$. 
	
	\begin{defn}An $(\FF,\TT)$-stopping time is an $\FF$-stopping time $\tau$ that can be written as ${\tau = \sum_{i=0}^n t_i \1_{A_i}}$ for some family $(A_i)_{i\in\{0,\cdots,n\}}$ of disjoint sets such that $\P(\cup_{i=0}^n A_i) = 1$ and $A_i\in\cF_{t_i}$. 
	\end{defn}				
	
	Throughout this section, we define $B^{\TT,\tau}_i\coloneqq\{\tau>t_i\}=\cup_{j=i+1}^n A_j$ and $i(t;\TT)\coloneqq\max\{i:t_i\leq t\}$. When it is clear which $\TT$ and $\tau$ we are referring to, we drop the subscript and write $i(t)=i(t;\TT)$ and $B_i = B^{\TT,\tau}_i$. Note that ${\{\tau>t\}=\{\tau>t_{i(t;\TT)}\}}$ for all $t\geq0$.
	
	For the first results in this section, we crucially need that the the filtration is constant between the points in $\TT$.
	
	\begin{condition}\label{cond:F tau}
		The pair $(\FF,\TT)$ satisfies Condition~\ref{cond:F tau} if $\cF_t = \cF_{t_{i(t;\TT)}}$ for all $t\geq0$.
	\end{condition}

	We first prove the existence of a proper solution associated to $(h_{EZ},U,\FF)$ where $U=(U_t)_{t\geq0}$ is given by $U_t=e^{-\gamma t}\1_{\{t<\tau\}}$. Here $\gamma>0$, and $\tau$ is an $(\FF,\TT)$-stopping time where $(\FF,\TT)$ satisfies Condition~\ref{cond:F tau}.
	\begin{prop}\label{prop:evaluating e^-gamma t 1[0,tau) finite valued}
		Let $\FF=(\cF_t)_{t\geq0}$ be a filtration and $\TT$ be an ordered set such that $(\FF,\TT)$ satisfies Condition~\ref{cond:F tau}. Let $\tau$ be a $(\FF,\TT)$-stopping time and define $U=(U_t)_{t\geq0}$ by $U_t=e^{-\gamma t}\1_{\{t<\tau\}}$. Then, there exists a proper solution $W=(W_t)_{t\geq0}$ associated to $(h_{EZ},U,\FF)$ such that
		\begin{equation}\label{eq:W bounded below by exponential until stopping time}
			W_t \geq \left(\frac{1}{\gamma\theta}\cEX[t]{e^{-\gamma t}- e^{-\gamma(t\vee\tau)}}\right)^\theta, \quad \text{for all }t\geq0.
		\end{equation}
	\end{prop}
	
	The proof of Proposition \ref{prop:evaluating e^-gamma t 1[0,tau) finite valued} relies on the following lemma.
	\begin{lemma}\label{lem:e^gamma tau sum form}
		Suppose that $\tau = \sum_{i=0}^n t_i \1_{A_i}$ is a $(\FF,\TT)$-stopping time. Then,
		\begin{equation}\label{eq:sum form of E[e^gamma]}
			\cEX[t]{e^{-\gamma t}- e^{-\gamma(t\vee\tau)}}=\1_{B_{i(t)}}(e^{-\gamma t}-e^{-\gamma t_{i(t)+1}})~+\!\!\!\sum_{j=i(t)+1}^{n-1}\!\!\cEX[t]{\1_{B_j}}\!\left(e^{-\gamma t_j}-e^{-\gamma t_{j+1}}\right)
		\end{equation}
	\end{lemma}
	\begin{proof}
		Using the definition of $\tau$ and $B_j$, the fact that $B_{i(t)}\in\cF_t$, and rearranging the telescoping sum gives
		\begin{align}
			\MoveEqLeft[4]\cEX[t]{e^{-\gamma t}- e^{-\gamma(t\vee\tau)}}
			\\
			~=&~~~\sum_{j=i(t)+1}^n (e^{-\gamma t}-e^{-\gamma t_j})\P(A_j|\cF_t)
			\\
			=&~~~e^{-\gamma t}\P(B_{i(t)}|\cF_t)-\sum_{j=i(t)+1}^n e^{-\gamma t_j}(\P(B_{j-1}|\cF_t)-\P(B_j|\cF_t))
			\\
			=&~~~\1_{B_{i(t)}}(e^{-\gamma t}-e^{-\gamma t_{i(t)+1}})+\!\!\sum_{j=i(t)+1}^{n-1}\!\!\cEX[t]{\1_{B_j}}\!\left(e^{-\gamma t_j}-e^{-\gamma t_{j+1}}\right).~~\qedhere
		\end{align}
	\end{proof}
	
	\begin{proof}[Proof of Proposition \ref{prop:evaluating e^-gamma t 1[0,tau) finite valued}]
		Define $W=(W_t)_{t\geq0}$ recursively backwards by $W_\infty=0$ $\as{\P}$ and, for $t<\infty$, $W_t= \1_{B_{i(t)}}w\big(t,\xi_{t_{i(t)+1}}\big)$, where
		\begin{align}\label{eq:W discounted indicator recursive defn}
			\qquad w(t,y)=\left(y^\frac{1}{\theta} + \frac{1}{\gamma\theta}\left(e^{-\gamma t}-e^{-\gamma t_{i(t)+1}}\right)\right)^\theta\quad \text{and}\quad\xi_{t_{i(t)+1}} = \cEX[t_{i(t)}]{W_{t_{i(t)+1}}}.
		\end{align}
		We will first show that $W$ is a solution associated to $(h_{EZ},U,\FF)$. We then show that \eqref{eq:W bounded below by exponential until stopping time} holds and $W$ is proper.

		First, note that $F_U(W)_\infty \coloneqq\lim_{t\to\infty}F_U(W)_t\leq\lim_{t\to\infty}\int_t^\infty e^{-\gamma s}(\frac{e^{\gamma\theta s}}{\gamma\theta})^\rho\dd s = 0 = W_\infty$. We will now show that $W_t=F_U(W)_t=\cEX[t]{\int_t^\infty U_s W_s^\rho \dd s}$ for all $t\geq0$ by backwards induction. For the inductive step, fix $k\in\{0,\cdots, n-1\}$ and assume that $W$ satisfies $W_{t_{k+1}}=F_U(W)_{t_{k+1}}$ and that $t_k\leq t< t_{k+1}$. 	By the definition of $W_{t_{k+1}}$, since $\1_{B_k}\1_{B_{k+1}}= \1_{B_{k+1}}$, and by Condition~\ref{cond:F tau}, $\cEX[t]{W_{t_{k+1}}}=\cEX[t]{\1_{B_k}W_{t_{k+1}}}=\1_{B_k}\cEX[t_k]{W_{t_{k+1}}}=\1_{B_k}\xi_{t_{k+1}}$. Hence, combining this with the inductive hypothesis yields
		\begin{align}
			F_U(W)_t =&~\cEX[t]{\int_t^{t_{k+1}}U_s W_s^\rho \dd s + W_{t_{k+1}}}
			\\
			=&~\1_{B_k}\left(\int_t^{t_{k+1}} e^{-\gamma s} (w(s, \xi_{t_{k+1}}))^\rho \dd s + \xi_{t_{k+1}}\right).
		\end{align}
		If $\omega\in B_k^c$, then clearly $W_t(\omega)=0=F_U(W)_t(\omega)$. Assume instead that $\omega\in B_k$. Then, $\xi\coloneqq\xi_{t_{k+1}}(\omega)$ is known. Since $\lim_{t\nearrow t_{k+1}}w(t,\xi)=\xi$ and $\frac{\partial w}{\partial t}(t,\xi) =  - e^{-\gamma t}(w(t,\xi))^\rho$ for $t_k\leq t<t_{k+1}$, integrating yields $w(t,\xi) = \int_t^{t_{k+1}} e^{-\gamma s} (w(s, \xi))^\rho \dd s + \xi$. In particular,
		$$W_t(\omega) = w(t,\xi) = \int_t^{t_{k+1}} e^{-\gamma s} (w(s, \xi))^\rho \dd s + \xi = F_U(W)_t(\omega).$$
		Consequently, $W_t = F_U(W)_t$ for $t_k\leq t \leq t_{k+1}$, and hence for all $t\geq0$ by induction.
		
		We now show that $W=(W_t)_{t\geq0}$ defined in \eqref{eq:W discounted indicator recursive defn} is proper by proving the following statement recursively backwards for $k\in\{0,\cdots, n-1\}$:
		If $t_k\leq t<t_{k+1}$, then
		\begin{equation}\label{eq:induction inequality W}
			W_t^\frac{1}{\theta}\geq\frac{\1_{B_k}}{\gamma\theta} \left(\sum_{j=k+1}^{n-1}\cEX[t]{\1_{B_j}}\left(e^{-\gamma t_j}-e^{-\gamma t_{j+1}}\right) + e^{-\gamma t}-e^{-\gamma t_{k+1}}\right).
		\end{equation}
		Here, we define $\sum_n^{n-1}a=0$ for arbitrary $a\in\R$. Hence, the statement holds true for $k=n-1$ by the definition of $W$ in \eqref{eq:W discounted indicator recursive defn}. For the induction step, assume that it holds true for $k+1$ and let $t_k\leq t<t_{k+1}$ so that $i(t)=k$. Using the definition of $W$ given in \eqref{eq:W discounted indicator recursive defn}, Jensen's inequality, the inductive hypothesis and the fact that  $\1_{B_i}\cEX[t_i]{\1_{B_j}}= \cEX[t_i]{\1_{B_i}\1_{B_j}} = \cEX[t_i]{\1_{B_j}}$ for $i\leq j$,
		\begin{align}
			W_t^\frac{1}{\theta}~={}&~\1_{B_k}\left(\left(\cEX[t]{ W_{t_{k+1}}}\right)^{\frac{1}{\theta}} + \frac{e^{-\gamma t}-e^{-\gamma t_{k+1}}}{\gamma\theta}\right)
			\\
			\geq{}&~\frac{\cEX[t]{\1_{B_{k+1}}}}{\gamma\theta}\!\left(\sum_{j=k+2}^{n-1}\!\cEX[t]{\1_{B_j}}\!\left(e^{-\gamma t_j}-e^{-\gamma t_{j+1}}\right) + e^{-\gamma t_{k+1}}-e^{-\gamma t_{k+2}}\right)
			\\
			&~\quad+ \1_{B_k}\frac{e^{-\gamma t}-e^{-\gamma t_{k+1}}}{\gamma\theta}
			\\
			={}&~ \frac{1}{\gamma\theta} \sum_{j=k+1}^{n-1}\cEX[t]{\1_{B_j}}\left(e^{-\gamma t_j}-e^{-\gamma t_{j+1}}\right) + \frac{\1_{B_k}}{\gamma\theta} \left(e^{-\gamma t}-e^{-\gamma t_{k+1}}\right).
		\end{align}
		Hence, \eqref{eq:induction inequality W} holds for all $t$ for which $t_k \leq t < t_{k+1}$, and hence by induction for all $t< t_n=\infty$.
		
		Since the right hand side of \eqref{eq:induction inequality W} is equal to $\frac{1}{\gamma\theta}\cEX[t]{e^{-\gamma t}- e^{-\gamma(t\vee\tau)}}$ by Lemma \ref{lem:e^gamma tau sum form}, Equation \eqref{eq:W bounded below by exponential until stopping time} holds and $W$ is proper.
	\end{proof}

	We now show that for any continuous filtration $\FF$ and $\FF$-stopping time $\tau$, we can  find a proper utility process associated to $U_t=e^{-\gamma t}\1_{\{t<\tau\}}$ by approximating $(\FF,\TT)$ by a monotone sequence of pairs $(\FF^n,\TT^n)$ satisfying Condition~B.
	\begin{lemma}\label{lem:filtration stopping time approx}
		Let $\FF=(\cF_t)_{t\geq0}$ be a continuous filtration and let $\tau$ be a $\FF$-stopping time. Let $\TT^n=\{k 2^{-n}: k =0,1,\cdots, n2^n\}\cup\{\infty\}$ and define $\FF^n=(\cF^n_t)_{t\geq0}$ by $\cF^n_t = \cF_{2^{-n}\lfloor 2^n t\rfloor\wedge n}$, for $t\geq0$. Then, $\tau_n = \1_{\{\tau\leq n\}}2^{-n}\lceil2^n\tau \rceil+\infty\1_{\{\tau>n\}}$ is a $(\FF^n,\TT^n)$-stopping time. Furthermore, $(\FF^n,\TT^n)$ satisfies Condition~\ref{cond:F tau} for each $n\in\N$, $\tau_n\searrow\tau$ and
		$\cF^n_t\nearrow\cF_t$ for all $t\geq0$.	
	\end{lemma}
	\begin{proof}
		Note that $\tau_n$ takes values in $\TT^n$ and $\{\tau_n \leq t\}=\{\tau\leq n\}\cap\{2^{-n}\lceil 2^n\tau\rceil\leq t\}=\{\tau\leq 2^{-n}\lfloor 2^n t\rfloor\wedge n\}\in\cF^n_t$,
		so that $\tau_n$ is a $(\FF^n,\TT^n)$-stopping time. 
		In addition, for each $n\in\N$ and $t\geq0$, 
		$t_{i(t;\TT_n)} = 2^{-n}\lfloor 2^n t\rfloor\wedge n$. Hence, $\cF^n_t=\cF_{t_{i(t;\TT_n)}}= \cF^n_{t_{i(t;\TT_n)}}$ so that $(\FF^n,\TT^n)$ satisfies Condition~\ref{cond:F tau}. 
		It is easily checked that $\tau_n \searrow \tau$ and $\cF^n_t\nearrow\cF_t$ for all $t\geq0$.
	\end{proof}
	
	To prove Proposition \ref{prop:proper solution discounted 1_[sigma,tau)}, we will need the following lemma, which is a variant of Hunt's Lemma and the Reverse Fatou Lemma.
	\begin{lemma}\label{lem:hunt fatou}
		Let $(\Omega, \cG, \P)$ be a probability space and $(X_n)_{n \in \NN}$ a sequence of random variables bounded in absolute value by an integrable random variable $Y$. Let $(\cG_n)_{n\in\N}$ be an increasing family of $\sigma$-algebras and $\cG_{\infty} \coloneqq \bigcup_{n = 1}^\infty \cG_n$. Then, $\limsup_{n \to \infty}\E\left[X_n\,\middle\vert\,\cG_n\right] \leq \E\left[\limsup_{n\to\infty}X_n\,\middle\vert\,\cG_\infty\right],~\as{\P}$
	\end{lemma}
	\begin{proof}
		Let $Z_m=\sup_{n\geq m}X_n\in L^1$. Then, $\E\left[X_n\,\middle\vert\,\cG_n\right]\leq\E\left[Z_m\,\middle\vert\,\cG_n\right]$ for $n\geq m$. Taking the $\limsup$ and using the $L^1$-Martingale Convergence Theorem yields
		\begin{equation}\label{eq:huntfatou1}
			\limsup_{n\to\infty}\E\left[X_n\,\middle\vert\,\cG_n\right]\leq\lim_{n\to\infty}\E\left[Z_m\,\middle\vert\,\cG_n\right]=\E\left[Z_m\,\middle\vert\,\cG_\infty\right].
		\end{equation}
		Furthermore, by the conditional version of the Reverse Fatou Lemma (and since $Z_m\leq Y$),
		\begin{equation}\label{eq:huntfatou2}
			\limsup_{m \to \infty}\E\left[Z_m\,\middle\vert\,\cG_\infty\right]\leq\cEX[\infty]{\lim_{m\to\infty}Z_m}=\E\left[\limsup_{n\to\infty}X_n\,\middle\vert\,\cG_\infty\right].
		\end{equation}
		Combining \eqref{eq:huntfatou1} and \eqref{eq:huntfatou2} yields the result.
	\end{proof}
	\begin{prop}\label{prop:evaluating e^-gamma t 1[0,tau)}
		Let $\FF=(\cF_t)_{t\geq0}$ be a continuous filtration, $\tau$ an $\FF$-stopping time and $\gamma>0$. Let $U=(U_t)_{t\geq0}$ be given by $U_t=e^{-\gamma t}\1_{\{t<\tau\}}$. Then, there exists a proper solution $W=(W_t)_{t\geq0}$ associated to $(h_{EZ},U,\FF)$ such that $W_t \geq \big(\frac{1}{\gamma\theta}\cEX[t]{e^{-\gamma t}- e^{-\gamma(t\vee\tau)}}\big)^\theta$ for all $t\geq0$.
	\end{prop}
	\begin{proof}
		By Lemma \ref{lem:filtration stopping time approx}, we may choose a sequence ${(\FF^n=(\cF^n_t)_{t\geq0},\TT^n)_{n\in\N}}$ such that $(\FF^n,\TT^n)$ satisfies Condition~\ref{cond:F tau} for each $n\in\N$ and a $(\FF^n,\TT^n)$-stopping time $\tau_n$ such that $\tau_n\searrow\tau$, and $\cF^n_t\nearrow\cF_t$ for $t\geq0$.
		Since, for each $n\in\N$, the pair $(\FF^n,\TT^n)$ satisfies the conditions of Proposition \ref{prop:evaluating e^-gamma t 1[0,tau) finite valued}, for $U^n = (U^n_t)_{t\geq0}$ defined by $U^n_t =e^{-\gamma t}\1_{\{t<\tau_n\}}$, there exists a proper solution $W^n=(W^n_t)_{t\geq0}$ associated to the triple $(h_{EZ},U^n,\FF^n)$ such that
		\begin{equation}\label{eq:W^n inequality}
			W^n_t\geq\left(\frac{1}{\gamma\theta}\E\left[e^{-\gamma t}- e^{-\gamma(t\vee\tau_n)}\thinspace \middle\vert\thinspace\cF^n_t\right]\right)^\theta.
		\end{equation}
		Since $W^n$ is a solution (and therefore c\`adl\`ag by Remark \ref{rem:solutions are utility processes}), for all bounded stopping times $\sigma\leq\tau$,
		\begin{equation}\label{eq:W^n solution eq}	
			W^n_\sigma = \E\left[\int_\sigma^\tau U^n_s (W^n_s)^\rho  \dd s + W^n_\tau\thinspace \middle\vert\thinspace \cF^n_\sigma\right].
		\end{equation}
		
		Consider $\ol u = (\ol u(t))_{t\geq0}$ defined by $\ol u(t) = e^{-\gamma t}$ for $t\geq0$. Then, by taking derivatives, one finds that $W^{\ol u} = (W^{\ol u}_t)_{t\geq0}$ defined by $W^{\ol u}_t=\frac{e^{-\gamma\theta t}}{\gamma^\theta\theta^\theta}$ is a solution associated to $(h_{EZ},\ol u)$ (and any filtration). Furthermore, as $J^{\ol u^\theta}_t = \int_t^\infty (\ol u(s))^\theta \dd s = \frac{e^{-\gamma\theta t}}{\gamma\theta}$, $W^{\ol u}\stackrel{\OO}{=}J^{\ol u^\theta}=(J^{\ol u^\theta}_t)_{t\geq0}$ and $W^{\ol u}$ is the maximal solution associated to $\ol u$ by Proposition \ref{prop:fixed point solution is maximal solution for O(Lambda)}. Therefore, since $U^n_t \leq \ol u(t)$ for $t\geq0$, it follows from Proposition \ref{prop:maximal solutions increasing in U} that $W^n_t\leq W^{\ol u}_t\leq W^{\ol u}_0 = \frac{1}{\gamma^\theta \theta^\theta}<\infty$ for all $t\geq0$ and $W^n$ is bounded. Similarly, $U_t (W^n_t)^\rho\leq \ol u(t)(W^{\ol u}_t)^\rho$ and $\E[\int_0^\infty \ol u(t)(W^{\ol u}_t)^\rho \dd t] = W^{\ol u}_0<\infty$.
		
		Define $W^*_t = \limsup_{n\to\infty}W^n_t$ for each $t\geq0$. We now show that $W^*=(W^*_t)_{t\geq0}$ is a subsolution associated to $(h_{EZ},U,\FF)$.
		Taking the $\limsup$ in \eqref{eq:W^n solution eq} and using Lemma \ref{lem:hunt fatou} gives
		\begin{align}\label{eq:W^* pre sub}\qquad \quad
			W^*_\sigma = \limsup_{n \to \infty}\E\left[\int_\sigma^\tau U^n_s (W^n_s)^\rho \dd s+ W^n_\tau\thinspace \middle\vert\thinspace \cF^n_\sigma\right] \leq \E\left[\int_\sigma^\tau U_s (W^*_s)^\rho  \dd s+ W^*_\tau\thinspace \middle\vert\thinspace \cF_\sigma\right].\qquad
		\end{align}
		Let $Y_t = W^*_t +\int_0^t U_s (W^*_s)^\rho  \dd s$. Then, by \eqref{eq:W^* pre sub}, $Y=(Y_t)_{t\geq0}$ is an optional strong submartingale. It is therefore l\`adl\`ag (see \cite[Theorem A1.4]{dellacherie1982}) and, by the strong submartingale property, $Y_\tau\leq \E\left[Y_{\tau+}\thinspace \middle\vert\thinspace \cF_\tau\right]$ for all stopping times $\tau$. Consequently, $W^*$ is l\`adl\`ag, and $$W^*_\tau = Y_\tau-\int_0^\tau U_s (W^*_s)^\rho  \dd s\leq \E\left[Y_{\tau+}\thinspace \middle\vert\thinspace \cF_\tau\right]-\int_0^\tau U_s (W^*_s)^\rho  \dd s = \E\left[W^*_{\tau+}\thinspace \middle\vert\thinspace \cF_\tau\right].$$ Thus, $W^*_\sigma \leq \E\left[\int_\sigma^\tau U_s (W^*_s)^\rho  \dd s+ W^*_{\tau+}\thinspace \middle\vert\thinspace \cF_\sigma\right]$.
		In addition, the transversality condition for subsolutions holds since $\limsup_{t\to\infty}\EX{W^*_{t+}}\leq \lim_{t\to\infty}W^{\ol u}_t=0$, so that $W^*$ is a subsolution for $(h_{EZ},U,\FF)$.
		
		Since $W^*$ is nonnegative and $\as{\P}$ bounded above by $W^{\ol u}_0<\infty$, it is ($L^1$-)bounded. Furthermore, by \eqref{eq:W^n inequality}, the fact that $\tau_n\geq\tau$ for all $n\in\N$ and the $L^1$-Martingale Convergence Theorem,
		\begin{align}
			W^*_t \geq&~ \lim_{n\to\infty}\left(\frac{1}{\gamma\theta}\E\left[e^{-\gamma t}- e^{-\gamma(t\vee\tau_n)}\thinspace \middle\vert\thinspace\cF^n_t\right]\right)^\theta
			\\
			\geq&~ \left(\frac{1}{\gamma\theta}\lim_{n\to\infty}\E\left[e^{-\gamma t}- e^{-\gamma(t\vee\tau)}\thinspace \middle\vert\thinspace\cF^n_t\right]\right)^\theta
			\\ \label{eq:W^* lower bound}
			=&~\left(\frac{1}{\gamma\theta}\E\left[e^{-\gamma t}- e^{-\gamma(t\vee\tau)}\thinspace \middle\vert\thinspace\cF_t\right]\right)^\theta.
		\end{align}
		Let $W^U=(W^U_t)_{t\geq0}$ be the \textit{maximal} solution associated to $(h_{EZ},U,\FF)$. Then, by Proposition \ref{prop:maximal existence U<K Lambda}, $W^U$ is the maximal $L^1$-bounded subsolution. Combining this with \eqref{eq:W^* lower bound} gives that $W^U$ is a proper solution and
		\begin{equation*}
			W^U_t \geq W^*_t \geq\left(\frac{1}{\gamma\theta}\E\left[e^{-\gamma t}-e^{-\gamma(t\vee\tau)}\thinspace \middle\vert\thinspace\cF_t\right]\right)^\theta,\quad \text{for all }t\geq0.\qedhere
		\end{equation*}	
	\end{proof}	
	
	Finally, we remove the assumption that $\sigma=0$ and prove Proposition \ref{prop:proper solution discounted 1_[sigma,tau)}.
	\begin{proof}[Proof of Proposition \ref{prop:proper solution discounted 1_[sigma,tau)}]
		Let $U_t=e^{-\gamma t}\1_{\{\sigma\leq t<\tau\}}$ and $\widehat{U}_t = e^{-\gamma t}\1_{\{t<\tau\}}$. Then, by Proposition \ref{prop:evaluating e^-gamma t 1[0,tau)}, there exists a proper solution $\widehat{W}$ associated to the pair $(h_{EZ},\widehat{U})$ such that $\widehat{W}_t \geq \big(\frac{1}{\gamma\theta}\cEX[t]{e^{-\gamma t}- e^{-\gamma(t\vee\tau)}}\big)^\theta$. First, consider the event
		$\{t\geq\sigma\}$. Then, $U_s=\widehat{U}_s$ for all $s\geq t$,  and hence the (unique) maximal solutions $W$ and $\widehat{W}$ associated to $U$ and $\widehat{U}$ coincide at $t$. Hence, 
		\begin{equation*}
			W_t \1_{\{t \geq \sigma\}} = \widehat{W}_t \1_{\{t \geq \sigma\}} \geq \big(\frac{1}{\gamma\theta}\cEX[t]{e^{-\gamma t}-e^{-\gamma(t\wedge\tau)}}\big)^\theta \1_{\{t \geq \sigma\}}.
		\end{equation*}
		Next, consider the event $\{t < \sigma\}$. Using that $W_\sigma = \widehat W_\sigma \geq \big(\frac{1}{\gamma\theta}\cEX[\sigma]{(e^{-\gamma \sigma}- e^{-\gamma\tau})}\big)^\theta$, Jensen's inequality and the tower property of conditional expectations give
		\begin{align*}
			W_t \1_{\{t < \sigma\}} &=\1_{\{t < \sigma\}} \cEX[t]{W_\sigma} \geq \1_{\{t < \sigma\}} \cEX[t]{\big(\tfrac{1}{\gamma\theta}\cEX[\sigma]{e^{-\gamma \sigma}- e^{-\gamma\tau}}\big)^\theta}\\
			&\geq\big(\tfrac{1}{\gamma\theta}\cEX[t]{e^{-\gamma \sigma}- e^{-\gamma\tau}}\big)^\theta\1_{\{t < \sigma\}}.  
		\end{align*}
		Combining the above inequalities yields \begin{equation*}
			W_t \geq \left(\frac{1}{\gamma\theta}\cEX[t]{e^{-\gamma (t\vee\sigma)}- e^{-\gamma(t\vee\tau)}}\right)^\theta. \qedhere
		\end{equation*}
	\end{proof}
	
\end{document}